\documentclass[aip,jcp,amsmath,amssymb,reprint,longbibliography,groupedaddress,noeprint]{revtex4-2}
\draft % marks overfull lines with a black rule on the right
\usepackage[utf8]{inputenc}
\usepackage{braket}
\usepackage{amsmath}
\usepackage{appendix}
\usepackage{amsfonts}
\usepackage{comment}
\usepackage{cancel}
\usepackage{bbold}
\usepackage{graphicx}
\usepackage{float}
\usepackage{MnSymbol}
\usepackage{mathtools}
\usepackage{tabularx}
\usepackage{array}
\usepackage{bm}
\usepackage{placeins}
\usepackage{natmove}
\usepackage{listings}

\newcolumntype{Y}{>{\centering\arraybackslash}X}

\usepackage[colorlinks, linkcolor=blue]{hyperref}
\usepackage[nameinlink, capitalise]{cleveref}
\Crefname{section}{Sec.}{Secs.}
\Crefname{appendix}{Appendix}{Appendices}

\newcommand{\nn}{\nonumber} 
\newcommand\numeq[1]%
{\stackrel{\scriptscriptstyle(\mkern-1.5mu#1\mkern-1.5mu)}{=}}

\newcommand{\tcb}[1]{} %turning off blue text

\newcommand\commin[1]{\iffalse #1 \fi}%inline comment

\newcommand{\B}{{\mathrm{b}}}
\newcommand{\SB}{{\mathrm{sb}}}

\newcommand{\mr}{\mathrm}
\newcommand{\mc}{\mathcal}
\newcommand{\don}{{\mathrm{d}}}
\newcommand{\acc}{{\mathrm{a}}}

\newcommand{\W}{\mathrm{W}}

\newcommand{\red}[1]{{\color{red}#1}}
\newcommand{\eu}{\mathrm{e}^}
\newcommand{\rmd}{\mathrm{d}}
\newcommand{\Tp}{t_\mathrm{p}}

\newcommand{\e}{\ensuremath{\,\mathrm{e}}}
\newcommand{\mat}[1]{\ensuremath{\mathsf{#1}}}
\newcommand{\g}{\mat{g}}
\newcommand{\Q}{\ensuremath{\mathrm{qc}}}
\newcommand{\C}{\ensuremath{\mathrm{cc}}}
\newcommand{\qc}{\ensuremath{\mathrm{qc}}}
\newcommand{\cc}{\ensuremath{\mathrm{cc}}}
\newcommand{\q}{\ensuremath{\mathrm{q}}}
\renewcommand{\b}{\ensuremath{\mathrm{b}}}
\renewcommand{\c}{\ensuremath{\mathrm{c}}}

\newcommand{\pr}{\ensuremath{\mathrm{p}}}
\renewcommand{\S}{\ensuremath{\mathrm{s}}}

\newcommand{\ii}{\ensuremath{\mathrm{i}}}
\newcommand{\rd}{\ensuremath{\mathrm{d}}}
\newcommand{\thalf}{\ensuremath{\tfrac{1}{2}}}

\newcommand{\kBT}{\ensuremath{k_\mathrm{B}T}}
\renewcommand{\Re}{\operatorname{Re}}
\renewcommand{\Im}{\operatorname{Im}}
\newcommand{\Tr}{\ensuremath{\mathrm{Tr}}}
\newcommand{\tr}{\ensuremath{\mathrm{Tr}}}

\DeclareMathOperator{\csch}{csch}
\DeclareMathOperator{\sech}{sech}

\begin{document}

% \title{Detailed balance in mixed quantum--classical dynamics: Ellipsoid mapping}
\title{On detailed balance in nonadiabatic dynamics: From spin spheres to equilibrium ellipsoids}
%\title{Nonadiabatic dynamics in thermal equilibrium}
\author{Graziano Amati}
\altaffiliation{These authors contributed equally}
\author{Johan E. Runeson}
\altaffiliation{These authors contributed equally}
\author{Jeremy O. Richardson}
\email{jeremy.richardson@phys.chem.ethz.ch}
\affiliation{Laboratory of Physical Chemistry, ETH Z\"urich, 8093 Z\"urich, Switzerland}

\date{\today}

\begin{abstract}
Trajectory-based methods that propagate classical nuclei on multiple quantum electronic states are often used to simulate nonadiabatic processes in the condensed phase. A long-standing problem of these methods is their lack of detailed balance, meaning that they do not conserve the equilibrium distribution. In this article, we investigate ideas for how to restore detailed balance in mixed quantum--classical systems by tailoring the previously proposed spin-mapping approach to thermal equilibrium. %Starting from an analysis of the spin-mapping approach, which treats the two-level system as a spin of magnitude $\sqrt{3}/2$, we consider two generalizations. 
We find that adapting the spin magnitude can recover the correct long-time populations but is insufficient to conserve the full equilibrium distribution. The latter can however be achieved by a more flexible mapping of the spin onto an ellipsoid, which is constructed to fulfill detailed balance for arbitrary potentials. This ellipsoid approach solves the problem of negative populations that has plagued previous mapping approaches and can therefore be applied also to strongly asymmetric and anharmonic systems. Because it conserves the thermal distribution, the method can also exploit efficient sampling schemes used in standard molecular dynamics, which drastically reduces the number of trajectories needed for convergence. %When applied to rate calculations,
The dynamics does however still have mean-field character, as is observed most clearly by evaluating reaction rates in the golden-rule limit. %, especially in the golden-rule limit.
This implies that although the ellipsoid mapping provides a rigorous framework, further work is required to find an accurate classical-trajectory approximation that captures more properties of the true quantum dynamics.
\end{abstract}

\maketitle

%%%%%%%%%%%%%%% Paper draft %%%%%%%%%%%%%%%%%%%

\section{Introduction}
% \begin{itemize}
%     % \item mixed quantum-classical at equilibrium: why we are interested in it -> linear response, rates, thermal transport (molecular junctions?)
%     % \item problem of detailed balance/conservation of thermal distribution
%     % \item fix properties we want (t=0, t=inf, conservation thermal distribution, correct zero coupling limit)
%     \item previous attempts
%     \item SM more accurate at high temperature -> starting point -> fix new properties
%     \item SUMMARY: generalize SM to fix those properties,  
% \end{itemize}

Detailed balance is a crucial concept in both classical and quantum statistical mechanics. It underpins the validity of many important microscopic relations, such as the fluctuation--dissipation theorem, which have lead to computationally powerful algorithms especially for classical molecular dynamics (MD).\cite{tuckerman_book} 
Classical MD provides an internally consistent treatment of detailed balance because Hamiltonian evolution conserves the classical Boltzmann distribution. % detailed balance is consistently treated by Hamiltonian evolution.
% The principle of detailed balance is a cornerstone of both classical and quantum statistical mechanics.
% Because it is consistently treated by Hamiltonian dynamics, % Thanks to this property being obeyed by Hamiltonian dynamics, % Because fully classical trajectories inherently respects it, 
% classical molecular dynamics (MD) has become a powerful computational technique.
In a similar way, the exact quantum-dynamical propagator commutes with the quantum Boltzmann operator, which proves that detailed balance should be a property of all quantum systems in thermal equilibrium.

However, when attempting to include quantum effects within classical-trajectory methods, detailed balance turns out to be a much more elusive property.\cite{bader1994correlation}
Within the adiabatic Born--Oppenheimer limit, it is now possible to respect quantum detailed balance (including zero-point energy and tunnelling) through path-integral methods such as 
% Within the adiabatic Born--Oppenheimer limit, where most modern theory development has focused on the inclusion of nuclear quantum effects, such as zero-point energy and tunnelling, it is now possible to properly respect quantum detailed balance through path integral-based methods such as 
centroid molecular dynamics (CMD),\cite{cao1993cmd} ring-polymer molecular dynamics (RPMD),\cite{craig2004rpmd} and path-integral Liouville dynamics.\cite{liu2014PILD} These have in many ways replaced previous methods that lacked this property, such as the linearized semiclassical-initial value representation (LSC-IVR).\cite{miller2001SCIVR}
The situation is, however, quite different when it comes to \emph{mixed quantum--classical} systems. The standard example of such a system is a molecule with classical nuclei and multiple quantum electronic states, but the classical part could as well be any low-frequency environment (not necessarily harmonic), and the quantum part could represent excitons, polaritons, high-frequency vibrations, and so on.
% When the quantum effects involve nonadiabatic transitions, however, there is still no simple trajectory-based method that is guaranteed to preserve the correct equilibrium distribution without breaking other reasonable limits (such as Rabi oscillations and golden-rule rates), even when the classical nuclear approximation is valid. % The same problem applies more generally to any mixed quantum--classical system, where the classical part could represent bath modes, electromagnetic field modes, etc., and the quantum part could represent spins, excitons, polaritons, and so on.
% In other words, predictive %realistic %accurate 
Despite significant effort, simulating nonadiabatic
dynamics of such systems while obeying detailed balance %in a system of quantum electrons and classical nuclei 
remains a crucial unsolved problem of statistical mechanics.

After decades of development, a variety of trajectory-based
methods have been proposed to simulate such systems. However, to our knowledge, none of these methods is guaranteed to preserve the correct equilibrium distribution without breaking other reasonable limits (such as recovering Rabi oscillations for an isolated quantum system), even when the classical nuclear approximation is valid.
%respect detailed balance in general.
% general and efficient strategy to simulate the dynamics of such systems is to use trajectory-based methods.
% Out of the most widely used approximations for mixed quantum--classical dynamics, % do not in general obey this property. For example, 
For example, the Ehrenfest approach (also known as mean-field theory) is known to severely violate detailed balance and can therefore not be used to describe relaxation
to thermal equilibrium.\cite{parandekar2006ehrenfest,runeson2022chimia}
Another prominent example, fewest-switches surface hopping, \cite{tully1990hopping} is known to recover detailed balance only in certain limits (small adiabatic splitting or large nonadiabatic coupling).% but not in general
\cite{parandekar2005mixed,schmidt2008SH} Similarly, the symmetric quasiclassical windowing approach (SQC)\cite{Miller2016Faraday} obeys detailed balance in the limit of small diabatic coupling \cite{Miller2015SQC} but not in general.
%is often found to be more accurate in practice, 
% but ultimately comes with no guarantees and is known to break detailed balance in several important cases \cite{schmidt2008SH}.
Although simple rate descriptions such as secular Redfield theory do obey detailed balance by construction,\cite{may_kuehn} these methods are only valid in certain parameter regimes (e.g.,\ weak system--bath coupling and Markovian dynamics). % and not applicable to full atomistic simulations. % and simple environments.

% For many of these processes, it is common to adopt a mixed quantum--classical description. 
% For instance, in a typical molecular simulation, classical nuclei evolve on quantized electronic states.
%Although in this paper we focus on the molecular situation of coupled electronic and nuclear degrees of freedom,
% However, the quantum--classical separation is useful also in many other types of systems, where the classical part could represent vibrations (also anharmonic ones), bath modes, electromagnetic field modes, etc., and the quantum part could represent spins, excitons, polaritons, and so on. 

% The theory of linear response offers a unified description of these seemingly disparate processes and has become a standard tool for numerical simulations. For example, in classical molecular dynamics one can study the relaxation of thermal fluctuations through an ensemble of trajectories on an effective Born--Oppenheimer potential. However, although this type of dynamics provides a good approximation to quantum dynamics in many situations, it is insufficient to describe processes that involve nonadiabatic transitions between multiple potential-energy surfaces, which is typically the case in many of the processes mentioned above.
%Mixed quantum-classical
% To go beyond the Born--Oppenheimer approximation, but still avoid a fully quantum simulation of both nuclei and electrons (which is practially impossible for most realistic systems), 

% What we want
The aim of this article is to investigate ways of constructing trajectory-based nonadiabatic dynamics which strictly obey detailed balance.
% In order to enable accurate and efficient simulations of complex nonadiabatic processes, it is therefore important to develop improved methods that strictly obey detailed balance. 
Note that the term \emph{detailed balance} is used to encompass several different concepts in the literature. In this paper, we say that a quantum--classical method obeys detailed balance if it initializes equilibrium systems in the correct quantum--classical Boltzmann distribution and preserves this distribution over time, so that the dynamics are microscopically reversible and time-translationally invariant. 
%and (ii) for systems with finite relaxation times, thermal correlation functions decay to the correct long-time limit. % to a product of equilibrium averages.
% This definition is aligned with common usage in the theory of the condensed phase. 
Apart from being necessary for internal consistency, detailed balance also enables a powerful arsenal of sampling tools developed for classical MD, which are currently not formally applicable to the conventional nonadiabatic techniques mentioned above.
Apart from this property, the quantum--classical dynamics should fulfill a few other relevant features in order to provide a realistic description. 
%Another reasonable requirement is that, 
In the limit of zero nuclear--electronic coupling, they should recover the correct result for an isolated quantum system (known as Rabi oscillations in the two-level case). Finally, the dynamics should reduce to classical MD on the ground-state Born--Oppenheimer potential in the adiabatic limit. %Finally, in the opposite nonadiabatic limit, the theory should recover the correct golden-rule behaviour of inter-state transfer rates.

In the present paper, we investigate a potential solution to this problem. % of classical nuclei.
%What we don't look for (at least not in this paper)
% Here we treat only classical nuclei, 
Here, we limit ourselves to situations where the nuclei can be treated classically,
although we note that because no known nonadiabatic extension to RPMD obeys all the required prescriptions above,\cite{richardson2013nrpmd,ananth2013mvrpmd,richardson2017vibronic,chowdhury2017coherent,miller3isomorphic2018} the present development is also likely to be relevant to tackle quantum nuclei with the path-integral formalism in future work.

% Previous attempts
%There have already been a number of attempts to go beyond standard methods like Ehrenfest and surface hopping.
%One type of approach that is particularly relevant for the theory of the present paper
%The theory of nonadiabatic dynamics that we use in the present paper is based on a
Our treatment is based on mapping the quantum subsystem onto a classical counterpart in order to treat both quantum and classical degrees of freedom on an equal footing. This idea has a long history, originally through the
%An important development within the theory of nonadiabatic dynamics is based on 
%mapping the quantum subsystem onto a classical counterpart, in order to treat both quantum and classical degrees of freedom on an equal footing. 
% A particularly prominent approach, known as the 
Meyer--Miller--Stock--Thoss (MMST) mapping, \cite{meyer1979nonadiabatic,stock1997mapping} which uses a set of harmonic oscillators as the mapping space. % ``mapping hamiltonian'' is fixed to a set of harmonic oscillators, 
% the approach 
This mapping has lead to a variety of classical-trajectory methods\cite{miller2009mapping,kelly2012mapping,hsieh2012FBTS,huo2012focus,cotton2013b,miller2016review,liu2016,saller2019jcp} which evolve the nuclei with a mean-field force, reminiscent of the Ehrenfest approach, but start from a different initial distribution and typically lead to higher accuracy.
% Within this framework, one can construct correlation functions from classical trajectories in many ways. \cite{huo2012focus,cotton2013b,miller2016review,liu2016,saller2019jcp}
% These evolve the nuclei with a mean-field force reminiscent of the Ehrenfest approach, but starting from a different initial distribution which typically leads to higher accuracy. %different mean-field force compared to Ehrenfest and typically lead to higher accuracy. 
However, these methods still break detailed balance, and in certain situations the populations (weights of nuclear forces) may even become negative, which means that the nuclei effectively move on inverted potentials and can lead to unphysical predictions.
% Among other proposed solutions, methods based on the numerical solution of the generalized quantum master equation (GQME) offer general improvements in the long-time relaxation properties, as the errors intrinsic to the long-time mapping dynamics are minimized by calculating short-lived memory kernels \cite{kelly2016master, pfalzgraff2019GQME}. Although these GQME-based approaches can be quite accurate for non extremes systems, they do not intrinstically provide a solution to the problem of detailed balance.
% A potential 
A possible solution proposed by Müller and Stock in the late 1990s is to modify a parameter that can be thought of as the zero-point energy (ZPE) of the oscillators, which decreases the likelihood of negative populations. \cite{Stock1999ZPE} They proposed a criterion for the optimal value of this parameter based on the long-time limit of the dynamics, or in other words, an attempt to recover detailed balance. Further generalizations of this parameter have recently been suggested,\cite{he2021commutator} but without addressing the issue of detailed balance. %In the present paper, we propose a strategy which is similar in spirit but, in addition, in our formalism all parameters are rigorously determined \emph{a priori}. % and do not need to be fitted or guessed \emph{a priori}. 

% Spin mapping
In this paper, we do not employ mapping to harmonic oscillators, but instead use a phase space for $N$-level systems known as the Stratonovich--Weyl (SW) representation, which generalizes the concept of a Wigner representation to systems with symmetry SU($N$). 
%discrete analog of the Wigner representation. 
Several of the methods originally developed for the MMST mapping have now been adapted to the SW framework and typically lead to improved accuracy. \cite{spinmap,runeson2020,mannouch2020paperI,mannouch2020paperII}
For two-level systems, the SW phase space is closely connected to the classical-vector model of a spin$-\thalf$ with radius $\tfrac{\sqrt{3}}{2}$, which is why the corresponding methods are referred to as ``spin mapping''. The spin-mapping equations of motion have the same form as in Ehrenfest and MMST mapping but start from a different (``spherical'') initial distribution and use a value of the zero-point energy parameter that is uniquely determined by the spin magnitude. (One can also construct a spherical distribution within the framework of MMST mapping by constraining the total population,\cite{liu2019,shalashilin2019CSS} although the special value of the ZPE parameter is less apparent in this approach.) Incidentally, the value used in spin mapping %is optimal in the sense that its value of the ZPE parameter 
provides the optimal high-temperature approximation to the long-time populations, as will be demonstrated in this paper. At low temperature, however, spin mapping is known to suffer from the same inverted-potential problem as other mapping methods. One way to understand this problem is that when the upper state is very high in energy compared to $\kBT$, the system should effectively be treated as a single-level system (the ground state) rather than a two-level system, and in such cases an SU(2) mapping is no longer appropriate. 
% \red{One of the reasons for the problems is that the SU(2) group is only appropriate if two electronic states are accessible.
% Problems tend to arise when one of the states is very high in energy, where the dynamics would be better described by classical mechanics on the lowest state.}

% This paper's contribution
In this paper, we show that the problem of detailed balance, as well as the inverted-potential problem, can be solved (at least in the two-level case) by generalizing the SW representation to a form that is more appropriate for equilibrium dynamics. % a form of correlation functions more appropriate than the ones considered in previous spin-mapping approaches. 
This leads to a new phase-space framework that takes into account which states are thermally accessible and preserves detailed balance by construction. In the new theory, one can think of the ``spin'' as evolving not on a sphere, but on an ellipsoid whose centre and shape can adapt along the trajectories. %We shall therefore refer to the theory as an ``ellipsoid mapping''. 
The ellipsoid dynamics preserve Rabi oscillations and, in the limit that the states are well-separated compared to $\kBT$, reduce to adiabatic dynamics on the lower state. In the opposite limit of states close in energy compared to $\kBT$, the theory reduces back to the original (spherical) spin mapping. 
Away from these two limits, the theory provides a gradual transition between two- and one-level systems. 

After deriving the theory in \cref{sec:theory}, we assess the accuracy of the ellipsoid mapping in \cref{sec:results} by computing thermal correlation functions for a spin--boson model, particularly in the strongly asymmetric regime where the standard mapping methods break down. Such correlation functions are %quilibrium dynamics is 
relevant to calculate a variety of experimental properties, from spectra to reaction rates, through their connection to linear-response theory. %can be understood in terms of dynamical processes in and near thermal equilibrium. 
%This is relevant for a variety of applications such as
% Since the method is constructed to have the correct short- and long-time limits, it is expected to be relevant for describing processes with multiple timescales from initial ultrafast dynamics to slow relaxation to equilibrium, such as
The present theoretical framework can therefore be applied to the study of a variety of physical processes and phenomena, such as
exciton transfer, vibronic spectroscopy, molecular junctions, heat transport, etc.\cite{KuboBook,Nitzan}

\section{Theory}\label{sec:theory}
% \tcb{Johan starts}
We start by considering a general two-level nonadiabatic system described by the Hamiltonian in the diabatic representation
\begin{equation}\label{eq:H2lx}
    \hat{H} = \frac{\hat{p}^2}{2m} + \hat{V}(\hat{x}), \quad \hat{V}(x)= \begin{pmatrix} V_1(x) & \Delta^*(x) \\ \Delta(x) & V_2(x) \end{pmatrix},
\end{equation}
where $V_1$ and $V_2$ are potential-energy surfaces and $\Delta$ is the coupling between the two levels. Typically $\hat{x}$ and $\hat{p}$ would represent the nuclear degrees of freedom in an electronic--nuclear problem, but the same model could be used to represent any two-state quantum subsystem in an environment. %, e.g. of phonons or light modes.
Although the present treatment is limited to two levels, we expect that it can be generalized to multiple levels in the future, similarly to the standard spin-mapping approach.\cite{runeson2020}

When the system is in thermal equilibrium, its expectation values are given by
\begin{equation}
    \braket{\hat{A}} = \Tr[\hat{\rho} \hat{A}] = \frac{1}{Z} \Tr[\eu{-\beta\hat{H}} \hat{A}] ,
\end{equation}
where $\beta=1/\kBT$, the partition function is $Z=\Tr[\eu{-\beta\hat{H}}]$, and the quantum-mechanical trace is taken over both subsystem and environmental degrees of freedom.
% \red{In thermal equilibrium,
% the expectation value of an operator is
% \begin{equation}
%     \braket{\hat{A}} = \Tr[\hat{\rho} \hat{A}] = \frac{1}{Z} \Tr[\eu{-\beta\hat{H}} \hat{A}] ,
% \end{equation}
% where the quantum partition function is $Z=\Tr[\eu{-\beta\hat{H}}]$.
% }
Many experimentally measurable properties of the system, such as spectra,\cite{bader1994correlation,mukamel1995book} rates,\cite{chandler1987greenbook} %\cite{ishizaki2005} thermal reaction rates,\cite{Yamamoto2002SCIVR,miller1983} %\cite{richardson2014} 
and scattering functions \cite{amati2019} can be expressed in terms of equilibrium time-correlation functions.
% Within the quantum--classical framework, t
The standard correlation function between two operators $\hat{A}$ and $\hat{B}$ is defined as 
\begin{align}
    C_{AB}(t) &= \frac{1}{Z}\Tr[\e^{-\beta \hat{H}}\hat{A}(0)\hat{B}(t)],\label{eq:CAB}\\
    \hat{B}(t) &= \e^{\ii \hat{H}t}\hat{B}(0)\e^{-\ii \hat{H}t},\nn
\end{align}
where  %$\beta=1/\kBT$, $Z=\Tr[\e^{-\beta \hat{H}}]$, and 
% the time evolution refers to fully quantum-mechanical dynamics, and
% the quantum-mechanical trace is taken over both nuclear and electronic degrees of freedom, 
throughout this paper, we use units for which $\hbar=1$. %Here, the time evolution refers to fully quantum-mechanical dynamics, %(there is no rigorous definition of mixed quantum--classical dynamics)
% followed by the mixed quantum--classical approximation when evaluating the trace.
%Here, the trace denotes the fully quantum-mechanical trace. At this point, we cannot replace the full trace by a mixed quantum--classical trace (except at $t=0$) because the corresponding time evolution is undefined. 
In general, thermal correlation functions depend on the ordering of the operators inside the trace. The ordering that is most closely related to linear-response theory and experimentally measurable quantities \cite{Zwanzig} is not the standard one above, but the Kubo-transformed %canonical (or Kubo-transformed) 
correlation function, defined by
\begin{equation} \label{eq:kuboAB}
K_{AB}(t) = \frac{1}{\beta Z}\int_0^\beta \rd \lambda\, \Tr[\e^{-(\beta-\lambda)\hat{H}}\hat{A}(0)\e^{-\lambda\hat{H}}\hat{B}(t)] .
%\equiv \langle \hat{A}(0);\hat{B}(t)\rangle_\Q.
\end{equation}
Both $C_{AB}$ and $K_{AB}$ contain the same information and either of them can be computed from the other through a simple relation between their Fourier transforms.\cite{craig2004rpmd} %\cite{KuehnBook} 
However, $K_{AB}$ is more closely related to classical correlation functions as they are both real and have the same symmetry  %invariant 
under time-reversal, in contrast to $C_{AB}$. Adiabatic path-integral methods such as CMD and RPMD have therefore focused on this quantity when including nuclear quantum effects. Even though their dynamics are fictitious, what makes these approximations to quantum dynamics particularly appealing is that they reproduce quantum statistics not only at $t=0$, but also at later times.  One therefore says that they conserve the Boltzmann distribution.\cite{Althorpe2021Matsubara}
This is a direct consequence of obeying detailed balance.

% Unfortunately, evaluating these functions in a condensed-phase environment with quantum mechanics is an exponentially hard problem.
Following the same spirit, we will in this paper present a theory to approximate $K_{AB}$ for nonadiabatic problems. % for any pair of operators $\hat{A}$ and $\hat{B}$ 
% (for now under the assumption of classical nuclei).
For now, we limit the discussion to classical nuclei, which is a reasonable assumption if $\kBT$ is large compared to the zero-point energies of the nuclear modes (but not necessarily compared to the electronic energy scales).
% \subsection{Problem definition}
%To introduce the problem of nonadiabatic thermal correlations,
% In this paper, we limit ourselves to two levels, but we expect that the
The assumption of classical nuclei means that we replace
% In this work, we study systems for which it is a valid approximation to 
$\hat{x}$ and $\hat{p}$ by classical phase-space variables $x$ and $p$, % evolving under Hamilton's equations of motion, 
while the electronic dynamics are still treated quantum-mechanically. 
% This leads to a mixed quantum--classical 
% This classical treatment implicitly assumes that $\kBT$ is large compared to the zero-point energy of the nuclear modes, but not necessarily compared to the electronic energy scales.
% Because the electronic degree of freedom is still treated quantum-mechanically, we need to define

To define equilibrium properties within this mixed quantum--classical framework, we introduce the quantum--classical density matrix\cite{mauri1993QC}
\begin{equation}
    \hat{\rho}(x,p) = \frac{1}{Z_\qc}\e^{-\beta \hat{H}(x,p)},
\end{equation}
and the partition function 
\begin{equation}\label{eq:Zqc}
Z_\qc = \Tr_\qc[\e^{-\beta \hat{H}(x,p)}],
\end{equation}
where the quantum--classical trace is to be understood as a classical phase-space integral over the nuclear (environmental) part and a quantum trace over the electronic (subsystem) part,
\begin{equation}
    \Tr_\qc[\hat{f}] = \int \rd x \rd p \,\Tr_\q[\hat{f}(x,p)].
\end{equation}
% and is a classical-nuclear approximation of the fully quantum-mechanical trace.
% The subscript Q (for ``quantum'') refers to the quantum-mechanical trace over the electronic states.
% Formally, one may think of this as the $\hbar\to0$ limit of a partially Wigner-transformed density matrix. %, which is commonly used in the context of mixed quantum--classical dynamics.
% \cite{nielsen2001QCbrackets,leaf}
Based on this prescription, expectation values %of operators 
are defined as
\begin{equation}\label{eq:Aqc}
    \langle \hat{A}\rangle_\Q = \Tr_\qc[\hat{\rho}\hat{A}] = \frac{1}{Z_\qc}\int \rd x \rd p \, \Tr_\q [\e^{-\beta \hat{H}(x,p)}\hat{A}(x,p)].
\end{equation}
Likewise, we define the quantum--classical limit of the Kubo-transformed correlation function at zero time, % inner product (pair correlation) between $\hat{A}$ and $\hat{B}$ as the quantum--classical zero-time value of the Kubo-transformed correlation function,
\begin{equation}\label{eq:ABqc}
    %\langle \hat{A},\hat{B}\rangle_\qc =
    K_{AB}^{(\mathrm{qc})}(0) = 
    \frac{1}{Z_\qc}\int \rd x\rd p\, \frac{1}{\beta}\int_0^\beta \rd\lambda \,\Tr_\q[\e^{-(\beta-\lambda)\hat{H}}\hat{A}\e^{-\lambda\hat{H}}\hat{B}],
\end{equation}
where the dependence on the nuclear variables has been suppressed for brevity. 
% \red{[check nielsen, kapral, ciccotti 2001 statistical mechanics of quantum-classical systems, and cite if relevant]}

The expressions in \cref{eq:Aqc,eq:ABqc} only report on statistical properties and it is not obvious how to generalize them to time-evolved quantities
without breaking time-translational invariance. For example, even the quantum--classical Liouville equations (QCLE), which has been used to derive many classical-trajectory methods, is known to break this property in general\cite{nielsen2001QCbrackets} (although it is exact for the spin--boson model).
% because $\hat{B}(t)$ does not have a clear quantum--classical limit. 
One strategy to define a quantum--classical limit for $\hat{B}(t)$ is to express not only the environment but also the subsystem in a classical phase-space picture, so that nuclear and electronic degrees of freedom are treated on the same footing. This is the philosophy of quasiclassical methods such as the MMST and spin mappings.
In \cref{sec:Wmapping,sec:optr}, we summarize the main features of spin mapping and demonstrate that it
tends to the wrong thermal distribution when the energy gap is large compared to $\kBT$,
%recovers detailed balance only for systems where the energy gap is small compared to $\kBT$.  
% For larger energy gaps, it is prone 
which makes it prone to
the same inverted-potential problems as the MMST mapping.
To overcome these problems, we will generalize the SW representation to the equilibrium case in \cref{sec:desiredproperties} % construct a different electronic phase space equipped with: (i) an equilibrium distribution that reproduces the expectation values in \cref{eq:Aqc,eq:ABqc}; and (ii) equations of motion that preserve this distribution and retain Rabi oscillations.
% In \cref{sec:desiredproperties}, we formulate the desired properties of such a phase space, which we then use to 
and use it to develop a new ellipsoid mapping method in \cref{sec:ell_isolated,sec:ell_mixed}.

\subsection{Summary of (spherical) spin mapping}\label{sec:Wmapping}
Most previous work on mixed quantum--classical dynamics has focused on calculating nonequilibrium correlation functions with factorized initial conditions,
% Previously, we have developed a mixed quantum-classical approach to calculate correlation functions of the type
\begin{equation}\label{eq:trAB}
    C_{AB}^{\rm b}(t) = \Tr[\hat{\rho}_{\rm b}\hat{A}(0)\hat{B}(t)],
\end{equation}
% i.e. without the thermal density operator.
where $\hat{\rho}_{\rm b}$ is a purely nuclear density. This density % In the classical-nuclear limit, $\rho_\b$ 
is typically represented % which can in principle be any distribution, but is typically chosen 
either as a Wigner function or as a classical Boltzmann distribution for the bath Hamiltonian (i.e., not the total Hamiltonian). The heart of the classical mapping technique is to define an analogous representation for the electronic operators. In this section, we summarize the main points of one such method, called spin mapping.\cite{spinmap} % which we will then extend to treat the special case of thermal correlation functions.
% Since the new theory of the present paper will extend upon this previous work, we start by summarizing its main points.

% Mixing quantum and classical dynamics is known to be a difficult problem plagued by mathematical inconsistensies [cite]. To derive a self-consistent theory, we want to introduce a phase-space representation for the quantum degrees of freedom as well as for the classical, so that both are treated on the same footing. For this purpose, there exists a rigorous phase-space theory for two- and higher-level systems that was pioneered by Stratonovich and Weyl. \cite{stratonovich1957distributions,Weyl1927} For two-level systems, the theory is closely related to the well-known equivalence to a spin-$\thalf$.
% The so-called Stratonovich--Weyl representation provides a discrete analog of the Wigner representation.

As is well known, a two-level system is isomorphic to a spin-$\thalf$ in a magnetic field, since the Hamiltonian is of the form
\begin{equation}\label{H2lexp}
    \hat{H} = H_0 \hat{\sigma}_0 + H_1 \hat{\sigma}_1 + H_2 \hat{\sigma}_2 + H_3 \hat{\sigma}_3 \equiv H_\mu \hat{\sigma}_\mu,
\end{equation}
where $\hat{\sigma}_\mu$ are the Pauli matrices (including $\hat{\sigma}_0$ as the $2\times 2$ identity) and repeated Greek indices imply summation from 0 to 3, whereas Latin indices will be used for 1 to 3. %As a notation, 
% In the following we reserve the sum over Latin indices to the subspace of the three Pauli matrices $i=1,2,3$. 
The explicit relations between \cref{eq:H2lx} and \cref{H2lexp} are $H_0 = \frac{p^2}{2m}+\thalf(V_1+V_2)$, $H_1=\Re\Delta$, $H_2=\Im\Delta$, $H_3 = \thalf(V_1 - V_2)$ and the corresponding ``magnetic field'' is the vector $\bm{H}=(H_1,H_2,H_3)$. \footnote{Note that some of the present expressions differ by a factor 2 from previous work which used a basis of $\hat{S}_i=\thalf\hat{\sigma}_i$; the form presented here is equivalent but leads to slightly neater equations.} % some of the present expressions differ the present notation differs from previous work\cite{runeson2019} by a factor 2, but the mehtod is equivalent. 
The operators $\hat{A}$ and $\hat{B}$ can be expanded in a similar way.

% \red{maybe give EOM here}

A convenient phase-space construction for the spin degree of freedom is provided by the so-called Stratonovich--Weyl representation, \cite{stratonovich1957distributions} which has recently gained attention \cite{tilma2016,rundle2019} after having been largely overlooked for many decades.\cite{brif1999phase} The SW representation can be thought of as the finite-level version % discrete 
of the more widely known Wigner representation of continuous degrees of freedom. % which is  often used for the nuclear degrees of freedom. 
In the SW representation, the Pauli operators are replaced by classical functions on the Bloch sphere according to the prescription
\begin{equation}\label{eq:oldmap}
 \begin{cases} \hat{\sigma}_0 \mapsto \sigma^{\rm W}_0 = 1 & \\ \hat{\sigma}_i\mapsto \sigma_i^{\rm W} = g_\W u_i & i=1,2,3, \end{cases}
\end{equation}
where  $\bm{u}=(u_1,u_2,u_3)=(\sin\theta\cos\varphi,\sin\theta\sin\varphi,\cos\theta)$ and $g_\W=\sqrt{3}$. Here, $\theta$ and $\varphi$ are the usual Bloch-sphere angles. Then $H_\W=H_\mu\sigma_\mu^\W$ and similarly for other operators. (Note that functions of the Pauli operators, such as the exponential, need to be expanded into a linear combination before this rule can be applied.)
The additional factor $g_\W$ guarantees that traces of products of operators are equal to the corresponding classical integrals,
\begin{equation}\label{eq:TrS_mn}
    \Tr_\q[\hat{\sigma}_\mu\hat{\sigma}_\nu]=2\delta_{\mu\nu}=\int \rd\bm{u}\, \sigma^{\rm W}_\mu \sigma^{\rm W}_\nu,
\end{equation}
where %$\Tr_\S$ is a trace over the subsystem and 
$\rd \bm{u} = \frac{1}{2\pi}\rd\varphi\rd\theta\,\sin\theta$ denotes integration over the unit sphere (normalized such that $\int \rd \bm{u}=\Tr_\q[\hat{\sigma}_0]=2$).
Note that the factor $g_\W$ also appears in the familiar expression for the magnitude $\tfrac{\sqrt{3}}{2}=\sqrt{S(S+1)}$ of a spin $S=\thalf$, hence the name ``spin mapping''. 

Because this formalism ensures that quantum traces are equal to classical integrals, \cref{eq:trAB} can be approximated by a classical correlation function,
\begin{equation}\label{eq:CAB_W}
    C_{AB}^{\rm b}(t) \approx \int \rd x \rd p \rd \bm{u} \,\rho_{\rm b}(x,p) A_\W(x,p,\bm{u})B_\W(x_t,p_t,\bm{u}_t),
\end{equation}
where the time evolution is given by
\begin{subequations}
\begin{align}
    \dot{x} &= p/m, \label{eq:xdotW} \\
    \dot{p} &= -\frac{\partial}{\partial x} H_\W, \label{eq:pdotW} \\
    \dot{\bm{u}} &= 2\bm{H}\times\bm{u},
\end{align}
\end{subequations}
and $H_\W = H_0(x) +g_\W \bm{H}(x)\cdot\bm{u}$.
\Cref{eq:CAB_W} is exact in the case of an uncoupled subsystem. In general, however, 
this independent-trajectory treatment is only exact at $t=0$ as it neglects higher-order terms in the Moyal series analogously to the LSC-IVR approximation (classical Wigner dynamics). It is therefore referred to as ``spin-LSC''.

% \red{
% maybe you want to give the formula more explicitly, e.g.
% \begin{align}
%     \int \rho_\mathrm{b}(x,p) \int du A(0) B(t)
% \end{align}
% }

In recent work, the SW representation for the electronic states (and its generalization to more than two states) has been employed to simulate various types of nonadiabatic dynamics initialized in nonequilibrium factorized states.\cite{spinmap,runeson2020,runeson2021,mannouch2020paperI, mannouch2020paperII,mannouch2022spectrum,runeson2022fmo,amati2022gqme} However, the theory is not yet in a suitable form for equilibrium correlation functions, as we will demonstrate in the following.

\subsubsection{Problems of spin mapping in equilibrium}
So far in \cref{sec:Wmapping}, we have focused on the case when the zero-time observable is a product state between the subsystem and the environment. To represent the thermal correlation function based on an initial density $\e^{-\beta \hat{H}}$ in \cref{eq:CAB} or \eqref{eq:kuboAB}, one might (naively) attempt to use the form $\frac 1 {Z_\W}\int\rd x \rd p \rd \bm{u}\, \eu{-\beta H_\W}A_\W B_\W(t)$, where the symbol `W' refers to the SW representation of subsystem operators [\cref{eq:oldmap}] and $Z_\W = \int\rd x \rd p \rd \bm{u}\, \eu{-\beta H_\W}$. However, this does not recover the correct initial distribution because, like Wigner transforms, SW can only represent traces of two spin operators correctly but not higher-order combinations, i.e., $\e^{-\beta H_\W}\neq[\e^{-\beta \hat{H}}]_\W=\sum_n \frac{1}{n!}(-\beta)^n[\hat{H}^n]_\W $. % because $H_\W^n\ne[\hat{H}^n]_\W$ for $n>2$.
Hence, this does not recover $K_{AB}(t)$ even for $t=0$.

To get the correct zero-time correlations, one could in principle use the Kubo-transformed operator $\frac{1}{\beta Z_\qc}\int_0^\beta \rd \lambda \e^{-(\beta-\lambda)\hat{H}}\hat{A}\e^{-\lambda \hat{H}}$ in place of $\hat\rho_{\rm b} \hat{A}$ in Eq.~\eqref{eq:trAB}. (This expression is easily evaluated since all operators have $2\times 2$ matrix representations within the classical-nuclear treatment). More details on this ``Kubo-transformed'' formulation of spin mapping can be found in \cref{app:SM_thermal}. %  (at least for simple systems where the Kubo transform can be computed analytically). 
% \red{Note that one cannot simply use $\eu{-\beta H}$ as the initial thermal distribution as this does not automatically give the correct expectation values.}
However, even though the initial values of thermal correlations are now corrected, the ensemble dynamics under $H_\W$ still do not preserve thermal expectation values. In the long-time limit (assuming ergodicity) the distribution would actually tend towards $\e^{-\beta H_\W}$ [see \cref{eq:SM_ij_LT}], which we have already concluded is incorrect. % initial values of thermal correlations are now corrected, the ensemble dynamics under $H_\W$ %the ensemble 
%does not generally preserve the initial distribution. Furthermore in the long-time limit, assuming ergodicity, the system relaxes according to the invariant distribution $\e^{-\beta H_\W}$, which we have already concluded is incorrect. 
As a consequence, spin-mapping time-evolved expectation values $\langle B(t)\rangle^\W$ [as defined in \cref{eq:SM_i_ij}] are not constant in time, %, just like in other mixed quantum--classical methods.\cite{mauri1993QC,nielsen2001QCbrackets}
in complete contradiction to the concept of equilibrium. %The failure to preserve the Boltzmann distribution is a long-standing problem in mixed quantum--classical dynamics \cite{mauri1993QC,nielsen2001QCbrackets} %also for other mappings 
% which, despite much effort, has not yet been solved for nonadiabatic problems, even in the limit of classical nuclei. % In other words, spin-mapping dynamics do not properly preserve the equilibrium distribution (just as other mappings). 

%Irrespective of the initial state,
A further issue is that spin mapping in its current form is known to suffer from negative populations, which violates the rule that the density matrix should be positive definite. %In extreme cases, this can lead to unphysical predictions from the ensemble average. 
On the level of individual trajectories, negative populations cause the nuclei to effectively evolve on inverted potentials, which can lead to unphysical dynamics %numerical problems 
at least for steep potentials. 
This problem is also present for quasiclassical methods based on MMST mapping (but notably not in the Ehrenfest approach).
%unless the offending trajectories are removed in postprocessing.\cite{miller+cotton?} %In \cref{sec:optr},

\subsection{Spherical mapping with an optimized radius}\label{sec:optr}
% \red{[all of the following could be moved to the next section (on the topic of how to get a  spin mapping for thermal correlation functions).
% You could start with the naive guess of just $\braket{\eu{-\beta H}A B}$
% and explain why that doesn't given correct t=0 result (instead of bringing up strong mixing ideas).
% Then offer the idea of the Kuboization of A - this gets right answer at t=0 but doesn't preserve distribution - instead it thermalizes to naive result given above.]}

% Here, we ask to what extent the classical spin analogy is useful also to describe thermal statistics at an inverse temperature $\beta=1/(\kBT)$. 

% In this section, we ask to what extent the idea of a classical spin is useful to describe thermal statistics for a quantum two-level systems at different temperatures. %This question is interesting no matter what the initial state is,
% In fact, no matter what the initial state is, %because most condensed-phase systems 
% Regardless of the initial state, the system
% will eventually relax to its own thermal equilibrium in the long-time limit, which is not necessarily equal to the correct quantum-classical equilibrium.
% With a simple example, we show that spin mapping in its current form is the optimal choice for high temperatures but that it needs to be modified for low temperatures. % and analyse how it can be generalized.
In this section, we demonstrate that populations predicted from spin mapping become negative when the energy separation of the two levels is large compared to $\kBT$, and subsequently discuss possible solutions.
According to ergodic theory, it is assumed that all nontrivial systems (with a non-zero coupling to a thermal environment) will relax to an equilibrium distribution regardless of their initial (nonequilibrium) state.
It is thus necessary to determine the relevant expectation values of this equilibrium distribution.
Our point is most simply illustrated by considering an isolated subsystem, so that we only have to consider traces over the electronic degrees of freedom.
%It is assumed that all nontrivial systems (with a weak but non-zero coupling to a thermal environment) will relax to this distribution regardless of their initial (nonequilibrium) state.

\begin{figure}
    \centering
    \includegraphics[width=\columnwidth]{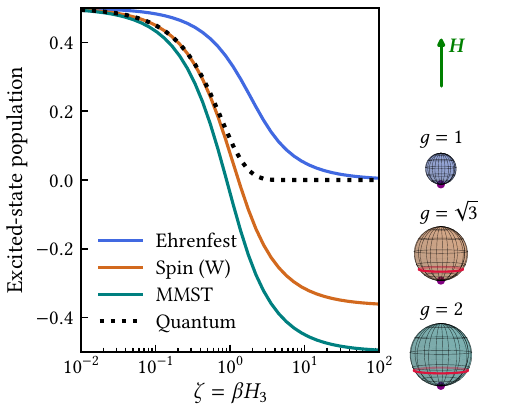}
    % Figure with two lines and two small cartoons of spins attached to them (showing that the classical spin points to the pole and misses the polar circle)
    \caption{Equilibrium population of the high-energy level within different quasiclassical mapping formalisms as a function of the dimensionless parameter $\zeta=\beta H_3$. Spin mapping is closer to the quantum result than Ehrenfest or MMST for high $T$ (low $\zeta$) but predicts negative populations for low $T$ (large $\zeta$). The reason is that for large $\zeta$, the spin distribution localizes to a single direction (purple dots on the spheres) antiparallel to the field (top right corner), which is different from the correct ground state (red circles). %if the field is directed upwards (as shown in the top right corner), then for large $\zeta$ the spin localizes to the purple dots on the spheres (with populations indicated by purple crosses in the plot), which is different from the correct ground state (indicated by red circles). 
    In this paper, we generalize spin mapping to reproduce the quantum populations for any $\zeta$. Figure adapted from Ref.~\onlinecite{runeson2022chimia}.} % alternative mapping that agrees with quantum equilibrium for any $\zeta$.}
    \label{fig:expval_z}
\end{figure}

Since it is always possible to find a basis in which ${H_1=H_2=0}$ (and because the theory is basis-independent), it suffices to consider the case ${\hat{H}=H_3\hat{\sigma}_3}$. Then $\langle\hat{\sigma}_1\rangle_\q=\langle\hat{\sigma}_2\rangle_\q=0$, while the remaining expectation value is
% $\hat{\sigma}_3\rangle$. Then the quantum expectation value is
\begin{equation}
    \langle \hat{\sigma}_3 \rangle_\q = \frac{\Tr_\q[\e^{-\beta H_3 \hat{\sigma}_3} \hat{\sigma}_3]}{\Tr_\q[\e^{-\beta H_3 \hat{\sigma}_3}]} = -\tanh\zeta, %\sim -\zeta+\tfrac{1}{3}\zeta^3 + \mathcal{O}(\zeta^5),
\end{equation}
where $\zeta \equiv \beta H_3$. This quantum result is to be compared with the corresponding classical expectation value that arises from the mapping $\hat{\sigma}_3 \mapsto \sigma_3 = g\cos\theta$, which is
\begin{equation}
    \langle \sigma_3 \rangle_\c = 
    \frac{\int \rd \bm{u}  \e^{-\zeta g \cos \theta}g \cos \theta}{ \int \rd \bm{u} \e^{-\zeta g \cos \theta} } = -g L(g\zeta), %\sim -\zeta + \tfrac{1}{5}\zeta^3 + \mathcal{O}(\zeta^5),
\end{equation}
where $L(x) \equiv \coth x - \frac{1}{x}$ is the Langevin function (a limiting case of the Brillouin function, which appears frequently in the paramagnetic theory of spins \cite{kittel}).
For $g=g_\W=\sqrt{3}$, the two expressions agree closely for small $\zeta$, as seen in \cref{fig:expval_z}, meaning that the classical phase-space average is valid if temperature is large or the energy separation between the two levels is small. A simple Taylor expansion shows that the error is of the order $\mathcal{O}(\zeta^3)$, a result unique to $g_\W=\sqrt{3}$, which is thus the optimal scaling in this regard.\footnote{For $N$ levels, the optimal choice is the scaling $g=\sqrt{N+1}$, but for three or more levels the error is of order $\mathcal{O}(\beta^2)$ instead of $\mathcal{O}(\beta^3)$. The two-level case is special because there the second-order contribution is identically zero due to the properties of the Pauli matrices.} 
For comparison, Ehrenfest (corresponding to a sphere with scaling $g=1$) and the original MMST mapping ($g=2$) both have an error of $\mathcal{O}(\zeta)$. 
For larger $\zeta$, however, the classical spin analogy breaks down %, since we are effectively dealing with a one-level system 
when the high-energy level becomes energetically inaccessible. Although it may seem that Ehrenfest improves again in this limit, one should keep in mind that the infinite-$\zeta$ case corresponds to an adiabatic system, where the direct use of Born--Oppenheimer MD would be preferable. %out would not require a nonadiabatic method, as Born--Oppenheimer molecular dynamics would be sufficient.

To improve the spin-mapping approach at larger values of $\zeta$, %we ask whether or not it is possible to modify this classical spin analogy in such a way that it reproduces the fundamental thermal properties of the quantum system for any value of $\zeta$. To match the expectation value in the example above, 
a simple solution would be to replace the factor $g_\W =\sqrt{3}$ with a $\zeta$-dependent scaling factor $g_\zeta$, defined as the unique (numerical) solution to $\tanh\zeta=g_\zeta L(g_\zeta \zeta)$. % $\langle \hat{\sigma}_3\rangle_\Q = = \langle g^* \cos\theta \rangle_\C$.
This procedure can be extended to also include the environment in the optimization of $g$, as described in \cref{app:optr}. A similar idea has been considered already by Müller and Stock in the context of the MMST mapping,\cite{Stock1999ZPE} where the so-called zero-point energy parameter $\gamma=g-1$ was optimized to obtain the correct long-time populations. However,
in both mapping formalisms, this value is dependent on the choice of basis and only fixes one of the three expectation values (except in the special case of an isolated system). Furthermore, this solution is not sufficient to preserve the equilibrium distribution (i.e., it does not give time-independent expectation values), as we will demonstrate in \cref{sec:results}.

It is also worth noting that by windowing the mapping variables, the SQC method predicts the exact long-time populations in this particular example.\cite{miller2015} %(although in its usual form, SQC does depend on the choice of basis).
Nonetheless, there are number of reasons why we do not employ the SQC approach in this work.
First, it has not been generalized to calculate thermal correlation functions.
Second, the results depend on the choice of the basis.
Finally, even when it predicts the correct (positive) populations on average, SQC can still suffer from the inverted-potential problem for individual trajectories.\cite{cotton2019trajZPE} % and is likewise prone to inverted forces.

%In addition, this solution would only be valid for expectation values, but not for correlations. % whereas the zero-time correlations $\langle \hat{\sigma}_i \hat{\sigma}_j \rangle_\C$ would not match the corresponding zero-time values of $K_{\sigma_i\sigma_j}(t)$ even for an isolated system.

% \red{[would be good to point out that it would not be so easy to do this within MMST mapping space as the thermal state would not be a singly-excited oscillator - cite NRPMD stuff and Ananth and Ananth+T.F.Miller.  This is thus a clear advantage of the smaller space available to spin mapping]}

% In a similar way, one can show that all second order expectation values $\langle \hat{\sigma}_i \hat{\sigma}_j \rangle_\Q$ agree with the classical ones only in the limit $\zeta\to 0$.
% This simple fix does not correct the second-order expressions $\langle \hat{\sigma}_i \hat{\sigma}_j \rangle_\Q$, which all would require a different value of $g$. A more serious problem is that 

% Optimal-radius idea, related to Müller and Stock's idea of picking gamma based on ergodicity (see \cref{fig:optz}).

% Why it is not enough 

\subsection{Desired properties of a phase-space representation for thermal equilibrium}\label{sec:desiredproperties}
In order to modify spin mapping such that it is more suitable to treat equilibrium correlation functions (as opposed to the nonequilibrium case of \cref{sec:Wmapping}), we start by analysing the defining properties of the self-dual or ``W'' Stratonovich--Weyl representation, which is the mathematical framework underlying the current spin mapping. %\cite{runeson2019,runeson2020} 
These are (for the subsystem degree of freedom):\cite{brif1999phase}
\begin{subequations}
\begin{enumerate}
\item Linearity: $\hat{A}\mapsto A(\bm{u})$ is a one-to-one linear map;
\item Reality:
    \begin{equation}
    [\hat{A}^\dagger](\bm{u}) = [A(\bm{u})]^*;
    \end{equation}
\item Normalization:
    \begin{equation}
    \Tr_\q[\hat{A}] = \int \rd\bm{u}\, A(\bm{u});
    \end{equation}
\item Tracing:
    \begin{equation}
    \Tr_\q[\hat{A}\hat{B}] =
    \int \rd\bm{u} \, A(\bm{u}) B(\bm{u});
    \end{equation}
\item Covariance:
    \begin{equation}\label{eq:covariance}
    [\hat{U}(G^{-1})\hat{A}\hat{U}(G)](\bm{u}) = A(G \cdot \bm{u}),
    \end{equation}
    where $G\in \mathrm{SU}(2)$ is a linear rotation operator and $\hat{U}(G)$ its unitary representation. % $\bm{u} \mapsto G\cdot\bm{u}$, and $\hat{U}(G)$ is a unitary operator. % (see Appendix~A for an example).
\end{enumerate}
\end{subequations}
Of these, property 4 is the one that enables an exact phase-space expression for the correlation functions [Eq.~\eqref{eq:trAB}] at zero time (and 3 is just a special case where $\hat{B}$ is the identity). In other words, it connects the trace product of quantum operators, $(\hat{A},\hat{B})_\q \equiv \Tr_\q[\hat{A}\hat{B}]$, with the ``classical'' inner product of functions on a sphere, $(A,B)_\c \equiv \int \rd\bm{u}\,A(\bm{u})B(\bm{u})$. However, there is flexibility in the definition of the inner product and these are %these are just two examples of inner products and 
not necessarily the best choice for the thermal case. In fact, an inner product that is more closely related to the canonical (Kubo-transformed) correlation in Eq.~\eqref{eq:kuboAB} is
\begin{equation}
    \langle \hat{A},\hat{B}\rangle_\q \equiv \int_0^1 \rd \lambda \, \Tr_\q[\hat{\rho}^{1-\lambda} \hat{A}\hat{\rho}^{\lambda}\hat{B}],
\end{equation}
where $\hat{\rho}=\frac{1}{Z_\q}\e^{-\beta H_i \hat{\sigma}_i}$ and $Z_\q =\Tr_\q[\e^{-\beta H_i \hat{\sigma}_i}]$. 
The classical analog of this inner product is
\begin{equation}
    \langle A,B\rangle_\c \equiv \int \rd\bm{u}\, \rho(\bm{u}) A(\bm{u})B(\bm{u}),
\end{equation}
where $\rho(\bm{u})=\frac{1}{Z_\c}\e^{-\beta H_i\sigma_i(\bm{u})}$ and $Z_\c = \int \rd \bm{u}\,\e^{-\beta H_i\sigma_i(\bm{u})}$. %(Here and in the following, we use angular brackets for weighted inner products and round brackets for unweighted ones.)

To make the Stratonovich--Weyl representation more suitable for this equilibrium problem, we propose to modify properties 3 and 4 to:
\begin{enumerate}
\begin{subequations}
\item[$3^\prime$.] Preservation of averages:
    \begin{equation}
     \int \rd\bm{u} \, \rho(\bm{u}) A(\bm{u}) = \Tr_\q[\hat{\rho} \hat{A}];
    %\langle \hat{A}\rangle_\Q.
    \end{equation}
\item[$4^\prime$.] Preservation of inner products:
    \begin{equation}
     \int \rd\bm{u} \, \rho(\bm{u}) A(\bm{u}) B (\bm{u}) = \int_0^1 \rd \lambda \, \Tr_\q[\hat{\rho}^{1-\lambda} \hat{A}\hat{\rho}^{\lambda}\hat{B}].
    %\langle \hat{A};\hat{B}\rangle_\Q,
    \end{equation}
\end{subequations}
\end{enumerate}
Note that in the limit $\beta\to 0$, these reduce back to the original SW properties 3 and 4.
Again, $3^\prime$ is just a special case of $4^\prime$ when $\hat{B}$ is the identity. 
We leave property 5 %[Eq.~\eqref{eq:covariance}]
unchanged, since it is the key to defining dynamics in phase space.\footnote{Note that by setting $\hat{U}=\e^{-\ii \hat{H}t}$, where $\hat{H}$ is the Hamiltonian of the subsystem, property 5 says that quantum time evolution should correspond to a rotation in the phase space of $\bm{u}$.} 
The two modified properties are not obeyed by the standard mapping [\cref{eq:oldmap}].
The problem of finding a new mapping $\hat{\sigma}_\mu \mapsto \sigma_\mu(\bm{u})$ to replace \cref{eq:oldmap} in order to fulfill these generalized Stratonovich--Weyl properties is the topic of \cref{sec:ell_isolated} (for the case of an isolated subsystem) and \cref{sec:ell_mixed} (for a mixed quantum--classical system). 
% In addition,
% \begin{enumerate}
%     \item[$6$.] Preservation of the partition function:
%     \begin{equation}
%         \int \rd\bm{u} \, \e^{-\beta (H_i \sigma_i(\bm{u})+\tilde{H})} = Z_\q.
%     \end{equation} 
% \end{enumerate}

As a final remark, an alternative way to express the generalized conditions is in terms of the \emph{generating functions}
\begin{subequations}
\begin{align}
F_\q(\bm{a}) &= \ln \,\Tr_\q[\e^{-(\beta H_i - a_i)\hat{\sigma}_i}], \\ 
F_\c(\bm{a}) &= \ln \int \rd \bm{u} \,\e^{-(\beta H_i-a_i)\sigma_i(\bm{u})}.
\end{align}
\end{subequations}
Then properties $3^\prime$ and $4^\prime$ are equivalent to matching the first two cumulants,
\begin{subequations}\label{eq:Fmoments}
\begin{align}
    \left. \left( \frac{\partial}{\partial a_i} F_\q\right)\right|_{\bm a \to \bm 0} &= \left. \left( \frac{\partial}{\partial a_i} F_\c\right)\right|_{\bm a \to \bm 0}, \\
    %\langle \hat{\sigma}_\mu; \hat{\sigma}_\nu\rangle_{\rm Q} - \langle \hat{\sigma}_\mu\rangle\langle \hat{\sigma}_\nu\rangle = 
    \left. \left( \frac{\partial^2}{\partial a_i\partial a_j} F_\q\right)\right|_{\bm{a} \to \bm 0} &= \left. \left( \frac{\partial^2}{\partial a_i\partial a_j} F_c\right)\right|_{\bm a \to \bm 0}.
 %= \langle \sigma_\mu \sigma_\nu\rangle_{\rm C}- \langle {\sigma}_\mu\rangle_{\rm C} \langle {\sigma}_\nu\rangle_{\rm C}
\end{align}
\end{subequations}
Finally, we will require $F_\q(\bm{0})=F_\c(\bm{0})$ in order to match the partition functions.

% [Say something about the covariance property]
%If, in addition, $\Tr[\hat{\rho}]=\int \rd \bm{u}\, \rho(\bm{u})$, then the mapping $\hat{A}\mapsto A(\bm{u})$ would provide a phase-space theory that preserves both the partition function and its first- and second-order expectation values. 
% [Discuss generating function here]
% Note that we do not require the partition functions to be equal.

% [Cite other papers about required for a statistical theory]

% Having demonstrated that no scalar $g$ can solve the problems, we instead propose the idea to promote this factor to a tensor, $g_{\mu\nu}$.

\subsection{Ellipsoid mapping for isolated subsystems}\label{sec:ell_isolated}
% To allow for wider applicability, we aim to extend the analogy even further and include second order thermal moments. It is easy to show that varying the spin magnitude by a scalar factor is not sufficient to match all of the second-order moments. 
We are now ready to construct a mapping $\hat{A}\mapsto A(\bm{u})$ that incorporates the properties described in the previous section. Since one can decompose any Hermitian operator as $\hat{A}=A_\mu \hat{\sigma}_\mu$, it suffices to define the mapping for each basis operator. As was made clear in \cref{sec:optr}, a simple scaling as in Eq.~\eqref{eq:oldmap} is not enough. % to match all the second-order moments. We therefore consider a more general modification. 
To proceed, it is helpful to consult a central theorem of quantum information theory, \cite{nielsenBook} which says that any trace-preserving quantum operation of the Bloch vector can be expressed as an \emph{affine map}
\begin{equation}
    \bm{u} \mapsto \g\bm{u} + \bm{c},
\end{equation}
where $\g$ is a real $3\times 3$ matrix (with elements $g_{ij}$) and $\bm{c}$ a translation vector.  % Since propagating a state to the equilibrium density is a \emph{bona fide} operation, % Since thermalization is an example of a quantum operation, 
% it is instructive to look for a solution of this form.
Further, the matrix $\g$ can be decomposed into a real orthogonal matrix (pure rotation) and a real symmetric matrix (deformation). Since thermal averages are computed from integrals over $\bm{u}$, they are unaffected by pure rotations. Therefore, we can take the matrix to be symmetric. % and for reasons to be explained shortly, we shall denote it $g$. 
Based on these considerations, we propose a mapping of the form
\begin{equation} \label{eq:anisotrop_mapping}
\hat{\sigma}_i \mapsto \sigma_i = g_{ij} u_j + c_j,
\end{equation}
% where $(u_0,u_1,u_2,u_3)=(1,\sin\theta\cos\varphi,\sin\theta\sin\varphi,\cos\theta)$ are basis functions on the unit sphere. 
where $g_{ij}=g_{ji}$. 
% $g_{\mu\nu}$ is a $4\times 4$ tensor of the form
% \begin{equation}\label{eq:ell_g_form}
% g = \begin{pmatrix}
% 1 & 0 & 0 & 0 \\ g_{10} & g_{11} & g_{12} & g_{13} \\ g_{20} & g_{21} & g_{22} & g_{23} \\ g_{30} & g_{31} & g_{32} & g_{33}
% \end{pmatrix}.
% \end{equation}
% The first row is required to make the identity map to 1 (so that the theory is invariant to a global shift of all energy levels). 
Geometrically,
one can think of this mapping as a transformation of the unit sphere into an ellipsoid, where
the elements $c_i$ specify its centre and $g_{ij}$ its shape.
% the elements $g_{i0}$ of the first column specify its centre and the $3\times 3$ symmetric submatrix $g_{ij}$ its shape. (Roman indices range from 1 to 3, in contrast to Greek indices which also include 0.) 

For completeness, we map the identity to one, $\hat{\sigma}_0\mapsto 1$, as usual (so that the theory is invariant to a global shift of all energy levels). In order for the mapping to preserve the partition function (equivalent to the zeroth cumulant of the generating function), we additionally introduce a scalar energy parameter $\tilde{H}$ %(and define $\tilde{H}_0=H_0+\tilde{H}$) 
in the mapping Hamiltonian, 
\begin{subequations}\label{eq:Hmap}
\begin{align}
    % \hat{H}\mapsto H_0 + \tilde{H}_0 + H_i g_{i0} + H_i g_{ij}u_j,
    \hat{H}\mapsto H &= H_i \sigma_i +  H_0 + \tilde{H}
    \\& = H_i g_{ij} u_j + H_i c_i + H_0 + \tilde{H}.
\end{align}
\end{subequations}
If $\bm{u}$ is thought of as an effective spin direction, $\g$ is analogous to the anisotropy tensor used to describe effective spins in electron paramagnetic resonance (EPR). \cite{abragam_bleaney}
% which is needed to define a density that is consistent with the original density. %, in order preserve the partition function.
%such that the mapping preserves the partition function. 
% This tensor needs to fulfill the requirements
% \begin{equation}\label{eq:mom_problem1}
%     Z_\Q = Z_\C
% \end{equation}

In total, we have introduced 10 independent parameters through the quantities $\g$, $\bm{c}$, and $\tilde{H}$.
These are defined to fulfill the requirements 
\begin{subequations} \label{eq:mom_problems}
\begin{align} \label{eq:mom_problem0}
    Z_\q &\stackrel{!}{=} Z_\c, \\
        \label{eq:mom_problem1}
    \langle \hat{\sigma}_i\rangle_\q &\stackrel{!}{=} \langle \sigma_i\rangle_\c, \\
        \label{eq:mom_problem2}
    \langle \hat{\sigma}_i, \hat{\sigma}_j \rangle_\q &\stackrel{!}{=} \langle \sigma_i, \sigma_j \rangle_\c,
    % K^\Q_{ij}(0) = K^\C_{ij}(0)
\end{align}
\end{subequations}
which constitute a system of $10$ nonlinear equations (because the inner products are symmetric). % with $10$ unknowns (because we have chosen $g_{ij}=g_{ji}$). %At this point, it is not clear whether a solution even exists, or if it is unique.

\begin{figure}
    \centering
    \includegraphics{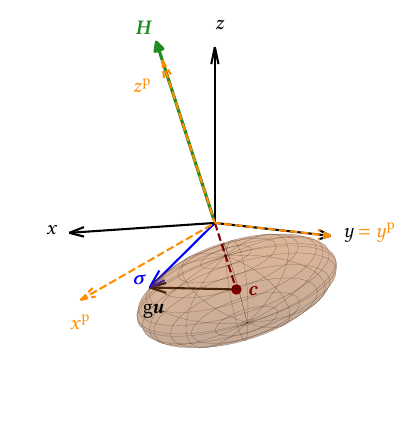}
    \caption{Visualization of the ellipsoid for an isolated two-level system. The ellipsoid is centred at $\bm{c}$ and $\bm{\sigma}=\g\bm{u}+\bm{c}$ is a vector on the surface (where $\bm{u}$ is a vector on the unit sphere). It is easiest to construct the ellipsoid in the principal-axis basis defined by the direction of the magnetic field, $\bm{H}$.}
    \label{fig:principal}
\end{figure}

This problem is most easily solved in a principal-axis basis where the $z$-axis is aligned with the magnetic field, as indicated in \cref{fig:principal}. (Rotating to this basis corresponds to diagonalizing $\hat{H}$.) % so that the magnetic field points along the $z$-axis. 
In this basis, the only non-zero expectation values are $\langle\hat{\sigma}_3\rangle_\q$ and $\langle \hat{\sigma}_i,\hat{\sigma}_j\rangle_\q$ for $i=j$. %To ensure the same for $\langle \sigma_i,\sigma_j\rangle_\c$, 
% This system is most easily solved in a principal-axis (diagonal) basis. To see that such a basis exists, note that $\langle \hat{\sigma}_i,\hat{\sigma}_j\rangle_\q$ is symmetric and therefore has a basis in which it is diagonal. 
To match the system of equations above, the same needs to be true for $\langle \sigma_i\rangle_\c$ and $\langle \sigma_i,\sigma_j\rangle_\c$. This is achieved if % this matrix is diagonal if %and only if 
$\g$ is chosen to be diagonal in the same basis, because then the Boltzmann factor does not depend on $\varphi$ and the integral over this variable is non-zero only for $i=j$. %exponential weighting $\e^{-\beta H_3^\pr g_{33}^\pr u_3^\pr}$. %Hence, there exists a % unique
% principal-axis basis.
%The symmetry of the problem also implies that there is a 
Further, the rotational symmetry around the $z$-axis also implies that
% The symmetry of the problem requires that one principal axis is oriented along $\bm{H}$ (which we choose to be $z$), around which there is a rotational symmetry, 
$\g_{11}^{\rm p}=\g_{22}^{\rm p}$. % (the shape of this ellipsoid is depicted in Fig.~\ref{fig:principal}). 
These considerations lead to the simpler form  % that the $g$-tensor % has the simpler form
\begin{equation}
c^{\rm p} = \begin{pmatrix}
0 \\ 0 \\ c_{3}^{\rm p} 
\end{pmatrix} ,
\qquad
\g^{\rm p} = \begin{pmatrix}
g_{11}^{\rm p} & 0 & 0 \\ 0 & g_{11}^{\rm p} & 0 \\
0 & 0 & g_{33}^{\rm p}
\end{pmatrix},
%  g^{\rm p} = \begin{pmatrix}
% 1 & 0 & 0 & 0 \\ 
% 0 & g_{11}^{\rm p} & 0 & 0 \\
% 0 & 0 & g_{11}^{\rm p} & 0 \\
% g_{30}^{\rm p} & 0 & 0 & g_{33}^{\rm p}
% \end{pmatrix},
\end{equation}
where `p' refers to the principal-axis basis. The transformation step to and from the original basis is given in \cref{app:principal}. There, we also show that in the principal-axis basis, the system of equations~\eqref{eq:mom_problems} reduces to a single nonlinear equation that is easily solved numerically and uniquely determines all the remaining parameters ($\tilde{H}$, $c^\mathrm{p}_3$, $g_{11}^\mathrm{p}$ and $g_{33}^\mathrm{p}$). The solution is then converted back to the original basis to perform the simulation. %calculate the averages.

\begin{figure*}
    \centering
    \includegraphics{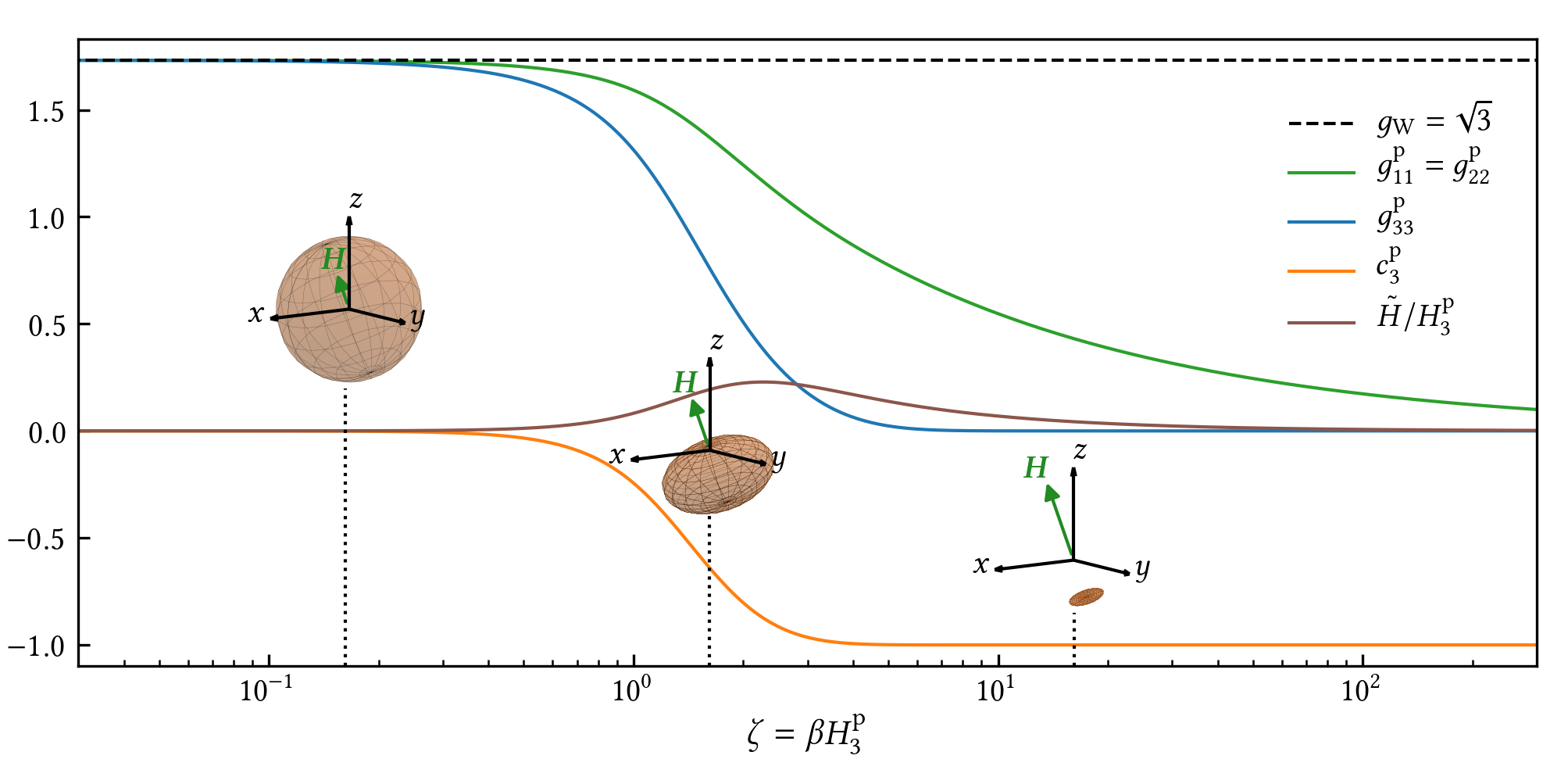}
    \caption{Ellipsoid-mapping parameters in the principal-axis basis for $\zeta>0$. In the limit of high temperature or low energy separation (left side), the ellipsoid reduces back to the W-sphere, and in the opposite limit (right side), it deforms into a point which represents the adiabatic ground state. The insets depict the gradual transition between these two limits. For $\zeta<0$, the signs of $c_3^\pr$ and $\tilde{H}/H_3^\pr$ are reversed, i.e., $\tilde{H}$ remains positive.}
    \label{fig:g_elements}
\end{figure*}

To analyse the solution, Fig.~\ref{fig:g_elements} shows all non-zero elements as a function of $\zeta=\beta H_3^{\rm p}$, which is the only relevant parameter for an isolated subsystem. The insets depict the ellipsoid for three different temperatures or field strengths (i.e., separation of the eigenenergies). First, one may note that for high $T$ or small energy separation, $c_3^\pr$ and $\tilde{H}$ vanish while $g_{33}^\pr$ and $g_{11}^\pr$ approach $\sqrt{3}$, meaning that we recover the original ``W-sphere'' spin mapping (\cref{sec:Wmapping}). In the opposite limit (low $T$ or large energy separation), the ellipsoid collapses to a point at distance 1 from the origin in the opposite direction from the magnetic field, and since $\tilde{H}$ is vanishingly small compared to $H_3^\pr$, this puts the spin vector into the adiabatic ground state. %which corresponds to the adiabatic ground state.
Between these limits, we observe that the spin vectors $\bm{\sigma}$ lie on an oblate ellipsoid (i.e., $g_{33}^\pr<g_{11}^\pr=g_{22}^\pr$), which changes smoothly along the transition between two-level and effectively one-level systems.
% Hence, populations cannot become negative once the surfaces are far apart, thereby closing the possibility of inverted potentials.
% Hence, the theory provides a gradual transition between two- and one-level systems.
The parameter $\tilde{H}$ does not influence the shape of the ellipsoid but can be thought of as a shift of the energy.

% As shown in \cref{fig:g_elements}, $c_3^\pr$ and $\tilde{H}$ vanish for small $\zeta$ while $g_{33}^\pr$ and $g_{11}^\pr$ approach $\sqrt{3}$, so that the dynamics in this limit is consistent with the spherical spin mapping. For large $\zeta$, the ellipsoid shrinks to a point with centre $c_3^\pr\to -1$, and since $\tilde{H}$ is vanishingly small compared to $H_3^\pr$, this limit corresponds to dynamics on the adiabatic ground state. 

% The latter corresponding to the special case $g = \mathrm{diag}(1,\sqrt{3},\sqrt{3},\sqrt{3})$. 

%In the following sections, we show how it is always possible to find a unique tensor $g$ that matches all quantum and classical moments up to order two. This correspondence will guarantee the correct statistical properties in the dynamics of electronic correlation functions. %Actually we look at cumulants, but moments sound friendlier

% Refer to appendix for full treatment

% including g00, solve moments zero, one, two

Before we couple the two-level system to the nuclear degrees of freedom, we make a brief comment about the spin dynamics. At this point, one may define the equations of motion either in terms of $\bm{\sigma}$ as 
\begin{subequations}
\begin{equation}\label{eq:ell_sigeq}
\dot{\bm{\sigma}} = 2\bm{H} \times \bm{\sigma},
\end{equation}
or equivalently in terms of $\bm{u}$ as
\begin{equation}\label{eq:ell_ueq}
\dot{\bm{u}} = 2\bm{H} \times \bm{u}.
\end{equation}
\end{subequations}
%%% Keep this comment
% \tcb{\begin{align*}
% \frac{\mr d}{\mr d t}(\g{\bm u}^\pr+\mb c) &= 2 \bm H\times (\g^\pr \mb u+\mb c)\nn\\
% \dot{\bm u} &= 2 (\g^\pr)^{-1}(\bm H\times (\g^\pr \mb u+\bm c))\nn\\
% \dot u_i &=  2 (\g^\pr_{ij})^{-1}\epsilon_{jkl}H_k (\g^\pr_{lm} u_m + c^\pr_l)  \nn\\
% \dot u_i &=  2 (\g^\pr_{ij})^{-1}\epsilon_{j3l}H_3 (\g^\pr_{lm} u_m + c^\pr_l)  \nn\\
% \dot u_i &=  2 \cancel{(\g^\pr_{ij})^{-1}}(\delta_{i2}\delta_{l1}-\delta_{i1}\delta_{l2})H_3 \cancel{\g^\pr_{ll}} u_l   \nn\\
% \dot u_i &=  2 \epsilon_{ijk} H_j u_l
% \end{align*}}
%%%
These equations describe the usual precession of the classical spin vector around the magnetic field.
The circular motions generated by both equations are equivalent because $\bm{H}$ is always aligned with one of the principal axes of the ellipsoid, around which there is a circular symmetry. Just like the spherical spin mapping, the dynamics are exact in the case of an isolated subsystem, such that
\begin{equation}
 \langle \hat{\sigma}_\mu(0), \hat{\sigma}_\nu(t) \rangle_\q = \langle \sigma_\mu(0), \sigma_\nu(t) \rangle_\c.
% K_{ij}(t) =  \frac{1}{\beta Z_\Q}\int_0^\beta \rd \lambda\, \tr[\e^{-(\beta-\lambda)\hat{H}}\hat{\sigma}_i(0)\e^{-\lambda\hat{H}}\hat{\sigma}_j(t)] =  \frac{1}{Z_\C} \int \dmu \, \e^{-\beta H_\mu \sigma_\mu} \sigma_i \sigma_j(t).
\end{equation}
In particular, the Boltzmann distribution is conserved, %expectation values are correctly initialized at $t=0$ and then preserved by the dynamics,
$\langle\sigma_i(t)\rangle_\c = \langle \sigma_i\rangle_\c = \langle\hat{\sigma}_i\rangle_\q =\langle \hat{\sigma}_i(t)\rangle_\q$.

\subsection{Ellipsoid mapping for mixed quantum-classical systems} \label{sec:ell_mixed}
We are now ready to include coupling to nuclear degrees of freedom (or in general to any type of classical environment). 
Our objective is to define a phase-space density for which
\begin{subequations} \label{eq:QCmom_problems}
\begin{align} \label{eq:QCmom_problem0}
    Z_\qc &\stackrel{!}{=} Z_\cc, \\
        \label{eq:QCmom_problem1}
    \langle \hat{\sigma}_i\rangle_\qc &\stackrel{!}{=} \langle \sigma_i\rangle_\cc, \\
        \label{eq:QCmom_problem2}
    \langle \hat{\sigma}_i, \hat{\sigma}_j \rangle_\qc &\stackrel{!}{=} \langle \sigma_i, \sigma_j \rangle_\cc,
\end{align}
\end{subequations}
and define dynamics that preserve equilibrium expectation values and reduce to the exact result in the isolated case. By the notation `cc' we mean `fully classical' expectation values
% approximate $\langle \hat{\sigma}_i(t)\rangle_\Q$ and $K_{\sigma_i\sigma_j}(t) = \langle \hat{\sigma}_i;\hat{\sigma}_j(t)\rangle_\Q$ by the fully classical integrals
\begin{align}
\langle f \rangle_\C &= \frac{1}{Z_\C}\int \rd x \rd p \int  \rd\bm{u} \, \e^{-\beta H(x,p,\bm{u})}f(x,p,\bm{u}), 
%  \langle \sigma_i(0);\sigma_j(t)\rangle_\C &= \frac{1}{Z_\C}\int \rd x \rd p  \int \rd\bm{u} \, \e^{-\beta H_\mu(x,p)\sigma_\mu}\sigma_i(0)\sigma_j(t), \label{eq:sig_ijC}
\end{align}
where
\begin{equation}
Z_\C = \int \rd x \rd p  \int \rd\bm{u} \, \e^{-\beta H(x,p,\bm{u})}.
\end{equation}

In order to make the fully classical and quantum--classical expressions agree at zero time, we need to further generalize the construction of the ellipsoid beyond that of the previous section, to take the additional nuclear degrees of freedom into account. To this end, we consider two different approaches. First, we attempt a \emph{global} construction in which $\g$, $\bm{c}$ and $\tilde{H}$ are independent of $x$. %As we shall see below, %this approach leads to (comparatively) simple dynamics, but it is on the other hand 
This is straightforward to implement for simple harmonic models where the nuclei can be integrated out analytically, but would be unpractical for more general potentials.
% it is only practical to find these mapping parameters for analytical potentials where the nuclei can be integrated out easily. % but on the other hand, this calculation only has to be carried out once at the start of the simulation. 
Second, we consider a \emph{local} construction with position-dependent $\g(x)$, $\bm{c}(x)$, and $\tilde{H}(x)$, which makes use of the solution from \cref{sec:ell_isolated} for each configuration $x$. This approach can be applied relatively easily to general potentials %, but has a somewhat more complicated %demanding
% dynamics because 
and uses an ellipsoid that changes shape and position along the trajectories.
% but requires recalculating the mapping parameters for each timestep.

Before we describe these two constructions in detail together with their equations of motion, let us make a list of properties that we wish the dynamics to obey:
\begin{enumerate}
\item Conserve energy, \cref{eq:Hmap};
\item Remain on the space $|\bm{u}|^2=1$; %, i.e. $\bm{\sigma}$ should remain on the ellipsoid.
\item Obey phase-space incompressibility,\cite{tuckerman_book} 
\begin{equation}
    \bm{\nabla}_{\bm{u}} \cdot \dot{\bm{u}} + \frac{\partial }{\partial x}\dot{x} + \frac{\partial}{\partial p}\dot{p} = 0,
\end{equation}
such that Liouville's theorem is valid;
\item Recover the exact Rabi oscillations in the limit of zero electron--nuclear coupling;
\item Reduce to evolution %dynamics 
on the adiabatic ground state in the limit of large $\beta|\bm{H}|$;
\item Visit the phase-space point $(x,p)$ with probability 
\begin{equation}
%    \frac{\Tr_\q[\e^{-\beta \hat{H}(x,p)}]}{\int \rd x\rd p \,\Tr_\q[\e^{-\beta \hat{H}(x,p)}]}.
    \frac{1}{Z_\qc} \Tr_\q[\e^{-\beta \hat{H}(x,p)}].
\end{equation}
\end{enumerate}
Together, these properties guarantee preservation of the equilibrium distribution and ensure physical dynamics in important limiting cases. % which can be used to make predictions about nonadiabatic transitions. 
In particular, property 2 is equivalent to the requirement that $\bm{\sigma}$ remains on the ellipsoid surface.
Along with Eqs.~\eqref{eq:QCmom_problems}, the classical equilibrium distribution is constructed to reproduce the correct quantum--classical results.
Property 6 additionally ensures that the expectation value of nuclear operators are also correct.

% Since the nuclear force is determined by $-\frac{\partial}{\partial x} (H_i\sigma_i+H_0+\tilde{H})$, an ellipsoid shrunk and shifted  $\sigma_i$ corresponds to the force of the lower adiabat.
%force $-\nabla_x (H_0+H_i\sigma_i)$ approaches $-\nabla_x(H_0-H_3)$, so if the ellipsoid is shrunk and shifted, the force must be on the lower adiabat. In contrast, for the spherical mapping, the nuclei will in general follow a mean-field force depending on the state populations.

\subsubsection{Global construction}
First, we attempt a global solution where $\g$ is independent of the nuclear variables. For this purpose, the solution in \cref{sec:ell_isolated} is no longer valid, but instead we propose to optimize $\g$ numerically as to solve \cref{eq:QCmom_problems} (a nonlinear multidimensional root problem of 10 unknowns). The additional nuclear integrals make this problem considerably harder to solve than the isolated case, but on the other hand, it only has to be solved once at the start of the simulation. For the special case of a spin--boson model, most of the bath degrees of freedom can be integrated out analytically using the reaction-coordinate representation of the Hamiltonian (see \cref{sec:spinboson,{app:QC_ave}}). %To simplify the calculations, we shall therefore concentrate on this model system when testing the perfomance of the method.

Compared to the isolated case, an important difference is that the local magnetic field is no longer aligned with any of the principal axes of the ellipsoid. This means that \cref{eq:ell_sigeq,eq:ell_ueq} are no longer equivalent, and one needs to construct the electronic equations of motion with more care. In fact, neither of these two options fulfills the requirements that we set up above: precession of $\bm{\sigma}$ about $\bm{H}$ leaves the surface of the ellipsoid, whereas precession of $\bm{u}$ about $\bm{H}$ (which does ensure $|\bm{u}|^2=1$) does not preserve the mapping Hamiltonian
% To construct the nuclear equations of motion, we write the mapping Hamiltonian on the form
\begin{equation}
\hat{H}(x,p) \mapsto H_0(x,p) + H_i(x) \sigma_i(\bm{u}) + \tilde{H}.
\end{equation}
In order to preserve energy as well as remain on the phase space, we instead propose to let $\bm{u}$ precess around the direction of %$\bm{H}^\intercal\g=
$\g \bm{H}$. Then requirements 1 and 2 are simultaneously fulfilled.  In addition, we note that % follows from the conservation of $|\bm{u}|^2$. 
the correct Rabi oscillations are obtained in the absence of electron--nuclear coupling if we pick the precession frequency such that
\begin{equation}\label{eq:ell_glob_Htilde}
 \dot{\bm{u}} = \frac{2}{g_\mathrm{eff}}(\g\bm{H})\times \bm{u}, 
 %2 \tilde{\bm{H}}\times \bm{u} ,
\end{equation}
where $g_{\rm eff}=\frac{|\g\bm{H}|}{|\bm{H}|}$. One may think of $g_{\rm eff}$ as an effective radius and the equations of motion correspond to Hamiltonian dynamics with conjugate variables $(\varphi,g_{\rm eff} \cos\theta)$.
The direction of precession is the same as for ``physical'' spins with an anisotropic $\g$-tensor in EPR, \cite{maryasov2012gtensor} but the precession frequencies differ by a factor $g_{\rm eff}$ due to the different physical significance of this quantity.

% \begin{equation}
%  \tilde{\bm{H}} = \frac{|\bm{H}|}{|\bm{H}^\intercal \mathbf{g}|} \bm{H}^\intercal \mathbf{g} .
% \end{equation}
% is chosen such that the precession frequency is always that of the local field,  which guarantees that Rabi oscillations are recovered in the absence of electron--nuclear coupling. 
% Begin with global construction for a spin-boson model, later local construction.

% \subsubsection{Dynamics with a global $g$ tensor}
% Conditions.
% Our proposal.
% Proof that it fulfills the conditions

% where
% \begin{equation}
%     H_0(x,p) =\frac{p^2}{2m} + \Tr_\q[\hat{V}(x)] .
% \end{equation}
Finally, it is clear that, along with the nuclear equations of motion in \cref{eq:xdotW,eq:pdotW},
% \begin{subequations}
% \begin{align}
% \dot{x} &= \frac{p}{m}, \\
% \dot{p} &= - \frac{\partial H_0}{\partial x}  - \frac{\partial H_i}{\partial x}\sigma_i(\bm{u}), %(g_{ij}u_j+c_i) 
% \label{eq:force_global}
% \end{align}
% \end{subequations}
the dynamics preserve the Hamiltonian and fulfill point 3 (using $\bm{\nabla}_{\bm{u}} \cdot \dot{\bm{u}}=0$). Note that the term $\tilde{H}$ %is independent of $x$ and 
does not influence the dynamics because it is independent of $x$ and may therefore be omitted within the global construction.

However, although the parameters of the global ellipsoid can be chosen to fulfill $\langle \sigma_\mu,\sigma_\nu\rangle_\cc=\langle \hat{\sigma}_\mu,\hat{\sigma}_\nu\rangle_\qc$, 
there is not enough flexibility to ensure that the equilibrium distribution over the nuclear phase space, $\frac{1}{Z_\cc} \int \rd \bm{u} \, \eu{-\beta H(x,p,\bm{u})}$, will recover the correct quantum--classical result according to property 6.
Thus, it only fulfills points 1--5 above but not point 6. 
This means that equilibrium averages will be correct only for electronic operators, but that in general nuclear expectation values will be incorrect.
%In addition, the global construction would be unpractical for anharmonic systems.
%As long as $g$ has the same limiting behaviour as in the isolated case, these reduce to adiabatic dynamics in the $|\bm{H}|\to\infty$ limit. 

\subsubsection{Local construction}\label{sec:local}
% A discomforting feature of the global construction is that it makes the nuclear force depend on the global shape of the potential. If, for example, the system is of a double-well type, then one would not expect the local force in one well to be influenced by the properties of the other well if they are far apart. % This is a consequence of the principle of locality which underlyes all known physical laws. 
To overcome the drawbacks of the global construction and fulfill all points 1--6, we consider an alternative local construction where the mapping parameters $\g(x)$, $\bm{c}(x)$, $\tilde{H}(x)$ are functions of nuclear configuration.
% \begin{equation}
% \hat{\sigma}_i \mapsto \sigma_i(x) = g_{ij}(x)u_i.
% \end{equation}
Effectively, this means that each $x$ defines a two-level system with a local potential matrix $\hat{V}(x)$. % for which we can reuse the solution from \cref{sec:isolated}.
To continue, define the local density $\hat{\rho}(x)=\frac{1}{Z_\q(x)}\e^{-\beta\hat{V}(x)}$, where $Z_\q(x)=\Tr_\q[\e^{-\beta\hat{V}(x)}]$ and the trace is taken only over the electronic subsystem (leaving $x$ unaffected). Next, define the local quantities
\begin{subequations}
\begin{align}
\langle \hat{\sigma}_i\rangle_{\q}(x) &= \Tr_\q[\hat{\rho}(x)\hat{\sigma}_i], \\
\langle \hat{\sigma}_i,\hat{\sigma}_j\rangle_{\q}(x) &= \int_0^1\rd\lambda \,\Tr_\q[\hat{\rho}^{1-\lambda}(x)\hat{\sigma}_i\hat{\rho}^\lambda(x)\hat{\sigma}_j].
% \langle \hat{\sigma}_i\rangle_{x,\Q} &= \frac{1}{Z_\Q(x)} \,\Tr_\q[\e^{-\beta\hat{V}(x)}\hat{\sigma}_i] \\
% \langle \hat{\sigma}_i,\hat{\sigma}_j\rangle_{x,\Q} &= \frac{1}{\beta Z_{\rm Q}(x)}\int_0^\beta \rd \lambda \, \Tr_\q[\e^{-(\beta-\lambda)\hat{V}(x)}\hat{\sigma}_i \e^{-\lambda \hat{V}(x)}\hat{\sigma}_j] \label{eq:sig_ijQx}
\end{align}
\end{subequations}
Analogously, define the phase-space density $\rho(x,\bm{u})=\frac{1}{Z_\c(x)}\e^{-\beta V(x,\bm{u})}$, where $Z_\c(x)=\int\rd\bm{u} \,\e^{-\beta V(x,\bm{u})}$ and the local potential is
\begin{equation}
    V(x,\bm{u}) = V_0(x) + H_i(x)\sigma_i(x,\bm{u}) + \tilde{H}(x), %+ H_i(x)\g_{ij}(x) u_j \\ + H_i(x) c_i(x) + \tilde{H}(x).
\end{equation}
with $V_0=\frac{1}{2}(V_1(x)+V_2(x))$. Finally, we define
\begin{subequations}
\begin{align}
\langle \sigma_i\rangle_{\c}(x) &= \int \rd \bm{u}\, \rho(x,\bm{u}) \sigma_i(x,\bm{u}), \\
\langle \sigma_i,\sigma_j \rangle_{\c}(x) &= \int \rd \bm{u} \,\rho(x,\bm{u})\sigma_i(x,\bm{u})\sigma_j(x,\bm{u}).
%  \langle \sigma_i\rangle_{x,\C} &= \frac{1}{Z_\C(x)}\int \rd\bm{u} \, \e^{-\beta H_k(x)\sigma_k(x)}\sigma_i(x) \\
%  \langle \sigma_i,\sigma_j\rangle_{x,\C} &= \frac{1}{Z_\C(x)}\int  \rd\bm{u} \, \e^{-\beta H_k(x)\sigma_k(x)}\sigma_i(x)\sigma_j(x), 
\end{align}
\end{subequations}
% where 
% \begin{align}
% && Z_\Q(x) &=\Tr_\q[\e^{-\beta \hat{V}(x)}], & Z_\C(x) &= \int \rd\bm{u}\, \e^{-H_k(x)\sigma_k(x)}. &&
% \end{align}
Then the solution from \cref{sec:ell_isolated} can be used to ensure that 
\begin{subequations}
\begin{align}
    Z_\q(x) &\stackrel{!}{=}Z_\c(x), \\
    \langle \hat{\sigma}_i\rangle_{\q}(x) &\stackrel{!}{=} \langle \sigma_i\rangle_{\c}(x), \\
    \langle \hat{\sigma}_i,\hat{\sigma}_j\rangle_{\q}(x) &\stackrel{!}{=} \langle \sigma_i,\sigma_j\rangle_{\c}(x).
\end{align}
\end{subequations}

In contrast to the global construction, the local ellipsoid will always be oriented with one principal axis along the magnetic field. Then the spin equation of motion, \cref{eq:ell_glob_Htilde}, reduces to the simpler \cref{eq:ell_ueq}. % so that there is no ambiguity about the spin part of the dynamics.
For the nuclear dynamics, the force needs to be modified compared to \cref{eq:pdotW} %\cref{eq:force_global} 
to take into account that $\sigma_i$ depend on $x$.
Unfortunately, there is not a unique way to determine the equations of motion.
Here, we suggest the following
\begin{subequations}
\begin{align}
\dot{\bm{u}}  &= 2\bm{H}\times \bm{u}, \\
\dot{x} &= \frac{p}{m}, \\
\dot{p} &= -\frac{\partial}{\partial x} V(x,\bm{u}).
\end{align}
\end{subequations}
%However, we emphasize that this is not a unique choice to fulfill the list of requirements.
By construction, these equations guarantee that not only electronic but also nuclear expectation values are correct and conserved, ensure incompressibility of phase space, recover Rabi oscillations for an uncoupled system, and reduce to single-surface dynamics on the lower adiabat in the large-$|\bm{H}|$ limit. 
Thus, this proposal obeys all 6 of the properties listed above.
%One may note that the compressibility is zero also in the space of $\bm{\sigma}$, i.e. $\nabla_{\bm{\sigma}}\cdot\dot{\bm{\sigma}}=0$ (which is not the case for the global construction).
Along a trajectory, the ellipsoid deforms and shifts according to the local potential-energy matrix. % such that the system remains in a thermal state. %such that the spin vectors $\bm{\sigma}$ remain in the correct equilibrium distribution. 
In regions of space where the excited state is high in energy compared to $\kBT$, the ellipsoid contracts so that the nuclei feel the ground-state force. When the states are close in energy, the ellipsoid becomes spherical and allows population transfer through precession of the spin vector.
Due to the reshaping of the ellipsoid, \cref{eq:ell_sigeq} is \emph{not} equivalent to \cref{eq:ell_ueq}, but instead $\bm{\sigma}$ has the more complicated equations of motion
\begin{equation}\label{eq:sigmadot}
    \dot{\bm{\sigma}} =\g\dot{\bm{u}}+\dot{\g}\bm{u}+\dot{\bm{c}}=  2\bm{H} \times \bm{\sigma} + \dot{x}\left(\frac{\partial \g}{\partial x}\g^{-1}(\bm\sigma-\bm{c}) + \frac{\partial \bm{c}}{\partial x}\right).
\end{equation}
% In the following sections we apply ellipsoid mapping to calculate the thermal dynamics of two-level nonadiabatic problems.
In practice, we sample and evolve the $\bm{u}$ variables and only convert to $\bm{\sigma}$ when calculating observables. The rotations of $\bm{u}$ were carried out in the standard way using quaternions. % which are related to the Cartesian mapping variables used in previous work.\cite{runeson2019} 
% The idea of an anisotropic mapping has been proposed already within the MMST-mapping framework ... \cite{he2021commutator}

\section{Model}\label{sec:spinboson}
In the last decades, the spin--boson model proved to be a valuable tool to investigate  decoherence and population transfer in nonadiabatic dynamics.\cite{garg1985, Mak1991spinboson, makri1995, Thoss2001hybrid} Despite its simplicity, this model has been shown to accurately describe a large class of relevant processes, from electron transfer to current flux in superconducting circuits. \cite{caldeira1983,xu1994,merkli2013, magazzu2018}
The out-of-equilibrium dynamics of the spin--boson model has been studied extensively in the literature.\cite{stockburger2004,mitra2005,boudjada2014,kundu2021} However, fewer studies have tackled the problem of thermal equilibrium correlation functions\cite{wang2006,montoya-castillo2017GQMEII,montoya-castillo2017PI,Mak1991spinboson,richardson2013nrpmd} as we do in the present work.

The Hamiltonian of the model can be split into three terms 
\begin{equation}
\hat H = \hat H_\S+  H_\B \hat\sigma_0 + \hat H_\SB,
\end{equation}
describing the uncoupled electronic subsystem, a harmonic bath of $F$ modes and the system--bath coupling interaction, respectively.  These terms are defined by
\begin{subequations}
\begin{align}
\hat H_\S &= \Delta \hat \sigma_1+\varepsilon \hat \sigma_3, \label{H_S} \\
 H_\B(x,p) &= \thalf\sum_{\alpha=1}^F\left( p_\alpha^2+m_\alpha\omega_\alpha^2x_\alpha^2\right),\label{H_B}\\
\hat H_{\SB}(x) &= \hat \sigma_3 \sum_{\alpha=1}^F c_\alpha x_\alpha. \label{H_SB}
\end{align}
\end{subequations}
Here, $\varepsilon$ denotes half the energy bias between the two electronic levels and $\Delta$ is the coupling constant between them. The frequencies $\omega_\alpha$ and the coupling constants $c_\alpha$ are related by the nuclear spectral density 
\begin{equation}\label{eq:JF}
J_F(\omega) = \frac \pi 2 \sum_{\alpha=1}^F \frac{c_\alpha^2}{m_\alpha\omega_\alpha}\delta(\omega-\omega_\alpha),
\end{equation}
and we set the masses of all nuclear modes to $m_\alpha=1$.
\Cref{eq:JF} is constructed as a discretization of a Debye spectrum 
\begin{equation}\label{eq:debye}
J(\omega)= \eta\frac{\omega\omega_c}{\omega^2+\omega_{\mathrm c}^2},
\end{equation}
% J[\[Omega]_] := \[Omega]*\[Omega]c*\[Eta]/(\[Omega]^2 + \[Omega]c^2)
% \[CapitalLambda] := Integrate[J[\[Omega]]/(Pi*\[Omega]), {\[Omega], 0, +Infinity}, Assumptions -> \[Omega]c In Reals && \[Omega]c > 0]
% \[CapitalLambda]
where $\eta$ denotes the strength of the system--bath coupling
(half the reorganization energy)
%(twice the reorganization energy) 
and $\omega_{\mathrm c}$ the characteristic frequency.
We observed satisfactory convergence of the correlation functions studied in this work with $F=200$ bath modes using a logarithmic discretization scheme. \cite{craig2007}

To justify a classical treatment of the nuclear statistics and dynamics, we
%use a high temperature compared to typical nuclear frequencies, by fixing
consider here the regime of high temperature
$\beta = 0.2$ and small nuclear frequencies $\omega_c = 1$.
As we will discuss in later sections, the agreement between full quantum and quantum--classical results validates such choice of parameters.
The electronic coupling constant is fixed to $\Delta = -1$. Numerically exact benchmark correlation functions are calculated using the hierarchical equation of motion (HEOM) method\cite{tanimura1989heom}. The standard thermal correlations obtained by HEOM are then converted to Kubo-form, as explained in  \cref{app:HEOM}.

For the spin--boson model it is particularly convenient to calculate the exact quantum--classical thermal averages defined in \cref{eq:Aqc,eq:ABqc}. %at the left-hand side of
%\cref{eq:QCmom_problem1,eq:QCmom_problem2}. 
Given that the electronic--nuclear coupling in \cref{H_SB} is a linear function of the  configurations, it is possible to integrate out analytically all nuclear modes in these phase-space averages, except for a one-dimensional reaction coordinate 
\begin{equation}\label{eq:RCy}
y = \sum_{\alpha=1}^F c_\alpha x_\alpha. 
\end{equation}
We refer to \cref{app:QC_ave} for details on these calculations.

\section{Results}\label{sec:results}
In this section, we test the global and local ellipsoid methods as well as the optimized sphere and compare their results with the original spin-LSC method (adapted for Kubo-transformed thermal correlation functions). 
For the spin--boson model described in \cref{sec:spinboson}, we consider a wide range of values of the electronic--nuclear coupling constant $\eta$ and of the energy bias $\varepsilon$. 
% Because of this, we will focus on this autocorrelation function. % in the following. 

\begin{figure*}[t!]
\centering
\includegraphics[width=7in]{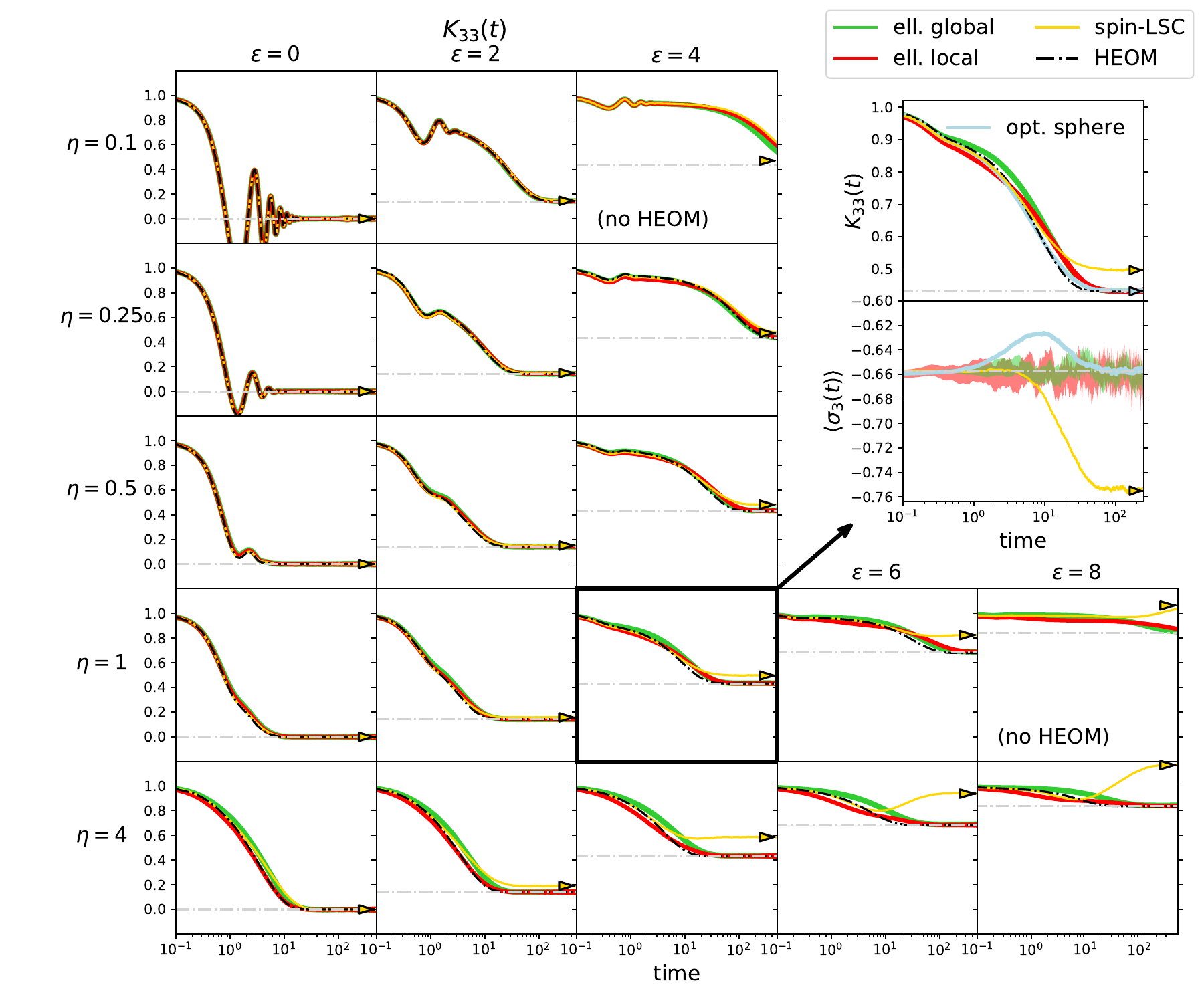}
\caption{
Thermal autocorrelation function of $\hat{\sigma}_3$ using the global and local ellipsoid (green and red, respectively), %(ell. global and ell. local), 
spin-LSC (yellow), and fully quantum HEOM results (black). %, and the generalization of spin mapping to an optimal sphere (light blue). 
Triangles indicate the theoretical predictions of the long-time limit of spin-LSC from \cref{eq:mix_KAB}, while the dash--dotted light grey lines show the correct quantum--classical long-time limits. %Numerically exact HEOM results are shown as black dash-dotted lines. % Mapping results are compared to the solutions of HEOM, shown here as black dot-dashed lines. 
Inset: expanded results for $\varepsilon=4$ and $\eta=1$, including the generalization of spin mapping to an optimized sphere (light blue). The lower panel of the inset shows the time evolution of  $\langle \sigma_3(t)\rangle$, which is expected to be constant provided detailed balance is obeyed. %for and optimzed formulation of thermal spherical spin mapping (presented in \cref{app:SM_thermal}). 
Shaded regions for the  global and local ellipsoid indicate the 95 \% confidence interval for the statistical average.
Note that time-averaging procedure [\cref{eq:t_ave}] was not used in this case, since it is not formally valid for the spherical methods, which do not obey detailed balance. In the case of the ellipsoid methods, it is valid, but we choose not to use it here in order to provide numerical evidence of this assertion.
}\label{fig:K33}
\end{figure*}

Of all correlation functions $K_{ij}(t)$, the hardest to reproduce with mapping techniques tends to be $K_{33}(t)$, especially for large $\varepsilon$. This is due to the fact that such correlation relaxes to nontrivial positive limit, which can be poorly captured as a consequence of the negative-population problem introduced in the discussion of \cref{fig:expval_z}. The time evolution of $K_{33}(t)$ is shown in \cref{fig:K33} as computed with the various methods for comparison with the HEOM benchmark. First, conventional spin-LSC (yellow lines) is found to predict incorrect long-time dynamics, especially for large $\varepsilon$. The details of how to apply spin-LSC to an equilibrium problem are given in \cref{app:SM_thermal}. There, we show that the long-time limit of this method is 
\begin{equation}\label{eq:mix_KAB}
\lim_{t\to+\infty}K^{\W}_{AB}(t) = \langle \hat A\rangle_{\qc} \langle  B\rangle_{\rm eq}^{\W},
\end{equation}
where
\begin{align}
\langle B\rangle_{\rm eq}^{\W} & \equiv \frac{\int \rmd x \rmd p \rmd \bm{u}\, \eu{-\beta H_{\W}} B}{\int \rmd x \rmd p \rmd \bm{u}\, \eu{-\beta H_{\W}}}.
\end{align}
The values predicted by \cref{eq:mix_KAB} are marked in \cref{fig:K33} by yellow triangles, which agree with the long-time limits of the yellow lines obtained from trajectory simulations. 
They are, however, \emph{not} equal to the correct long-time limit $\langle \hat A\rangle_{\qc}\langle \hat B\rangle_{\qc}$ (shown as grey dash--dotted lines, calculated as described in \cref{app:QC_ave}).
The error is observed to become more severe for higher values of $\varepsilon$ and $\eta$, and
for the most extreme system ($\varepsilon=8, \eta=4$), the long-time limit of spin-LSC is even found to be larger than 1. %the even relaxes to a limit above the initial value. 
This corresponds to an unphysical situation where the higher-energy state has a negative thermal population. 
We remark that the failure to relax to the correct limit (especially for strongly biased systems) is a problem not just for spin-LSC, but is present also in several other mapping techniques including Ehrenfest mean-field methods \cite{bastida2007,montoya-castillo2017GQMEII} and
approaches based on the MMST mapping, \cite{saller2019jcp,bellonzi2016assessment,coronado2001} including SQC. \cite{miller2015,zheng2017,tao2017} 

Next, we consider optimizing the spin radius in spin-LSC with respect to the long-time population (see \cref{app:optr} for details). This idea is closely connected to the approach by M\"{u}ller and Stock of optimizing the zero-point energy parameter in the MMST mapping.
The analogous strategy in the spin-mapping framework is to choose
the sphere radius $g$ such that $K_{33}^{g}(t)$ and $\langle \sigma_3(t)\rangle_\cc^{g}$ relax to the correct long--time limits. 
We include results for this approach in the inset on the top right corner of \cref{fig:K33}, for a representative system with $\varepsilon=4$ and $\eta = 1$.  The long-time limit of $K_{33}(t)$ is correct by construction. %The intermediate dynamics of this approach appear to be slighly more accurate than the two flavours of ellipsoid mapping. 
However, a caveat of the method is that the system is formally sampled from an out-of-equilibrium initial distribution (relative to the Hamiltonian which generates the dynamics). This implies that the time-dependent average $\langle \sigma_3(t)\rangle_\cc$ (shown with light blue line in the lower panel of the inset) is not preserved by the dynamics, which breaks the desired time-translational invariance. The optimized-sphere approach is also limited to fixing a single expectation value and will in general lead to an incorrect long-time limit of $\langle \sigma_1(t)\rangle_\cc$ (and hence also of $K_{31}(t)$ and $K_{13}(t)$, whilst $K_{21}(t)$ is anyway guaranteed to relax to the correct zero long-time limit by symmetry in this case). 

In contrast, the ellipsoid mapping approach recovers all the correct long-time limits by construction, as shown with green and red lines for the global and local constructions, respectively. Because these methods preserve their respective distributions and obey Liouville's theorem, it is possible to obtain converged results with a low number of trajectories by using the time-averaging procedure\cite{tuckerman_book} %equivalence between time and phase-space averages in ergodic systems \cite{tuckerman_book},
\begin{subequations}
\begin{align}\label{eq:t_ave}
    \langle A(0)B(t)\rangle_\cc &= \frac{1}{\tau_\mathrm{max}}\int_0^{\tau_\mathrm{max}} \rd \tau \,\langle A(\tau)B(t+\tau)\rangle_\cc
    %\int \rd \Gamma_0 \rho(\Gamma_0)A(\Gamma_0)B(\Gamma_\tau) \\ &=\int \rd \Gamma_t \rho(\Gamma_\tau)A(\Gamma_\tau)B(\Gamma_{t+\tau})
\end{align}
\end{subequations}
% where $\Gamma=(x,p,\bm{u})$.
For these models, we found that the order of $10^3$ trajectories of length $\tau_\mathrm{max}=10^3$ was sufficient for convergence. % (the results shown here used 3600 for local and 7200 for global).
Note, however, that while the ellipsoid methods relax by construction to the correct long-time limits, the intermediate dynamics are found to be less accurate than with spin-LSC or the optimized sphere for large $\eta$ and $\varepsilon$. In the models considered here, the global construction is found to overestimate and the local construction to underestimate the timescale of population transfer.

Finally, we remark that all methods are able to capture the transition from coherent to incoherent relaxation, which is seen to occur when increasing the electronic--nuclear coupling constant $\eta$.
This behaviour was first studied using real-time path-integral Monte Carlo,\cite{Mak1991spinboson} and it is interesting to note that it can be well described by simpler classical-trajectory methods. %Only for large $\eta$ and $\varepsilon$ does the intermediate dynamics deviate slighly from the HEOM results.

\subsection{Detailed balance in rate theory}
Having identified the local ellipsoid as a method that fulfills all the required properties in \cref{sec:ell_mixed}, we next discuss to what extent it is useful for calculating nonadiabatic rates. % and identify the main advantages and drawbacks of the approach.
In particular, we show that the ellipsoid approach formally obeys detailed balance as typically defined in a rate-theory framework,\cite{chandler1987greenbook}
%i.e., %Detailed balance in the framework of rate theory is the property that
%in equilibrium, the forward and backward reaction rates are equal, %can be formulated by requiring that the ``amount of population'' exchanged between the donor and acceptor is conserved, that is
\begin{equation}\label{eq:db_rates}
P_{\don}k_{\don\to\acc}=P_{\acc}k_{\acc\to\don}.
\end{equation}
For the sake of the present discussion, we identify the reactants with a donor (d) and the products with an acceptor (a). Then
$P_{\don} =\big\langle\ket{\don}\bra{\don}\big\rangle_\qc$ is the equilibrium probability of finding the system in the donor, and similarly $P_{\acc}=1-P_{\don}$.
These probabilities are multiplied by the rate constants $k_{\don\to\acc}$ and $k_{\acc\to\don}$ to obtain the forward and backward reaction rates, which according to \cref{eq:db_rates} must be equal in equilibrium.

First, we consider identifying the donor $\ket \don $ and acceptor $\ket \acc $ with the two electronic states, such that
\begin{equation}
\ket \don \bra \don = \tfrac{1}{2} (\hat\sigma_0+\hat\sigma_3), \hspace{7mm} \ket \acc \bra \acc = \tfrac{1}{2} (\hat\sigma_0-\hat\sigma_3).
\end{equation}
These definitions lead to  the thermal side--side correlation functions\cite{miller1983tromp}
\begin{subequations}
\begin{align}
K_{\don\acc}(t) &= \tfrac{1}{4} \big(K_{00}(t)-K_{03}(t)+K_{30}(t) -K_{33}(t)\big)
\\
K_{\acc\don}(t) &= \tfrac{1}{4} \big(K_{00}(t)+K_{03}(t)-K_{30}(t) -K_{33}(t)\big).
\end{align}
\end{subequations}
Based on these, %side--side correlation functions, %\cite{miller1983tromp}
the rate constant for the transition between $\ket \don$ and $\ket \acc$ is defined by \cite{chandler1987greenbook} % chapt 8.3 
%\begin{subequations}
\begin{align}
k_{\don\to\acc}^{(\mr s)} &= \frac {1}{P_\don}\lim_{t\to \Tp}  \dot K_{\don\acc}(t), \label{eq:k_da} %\\
%P_{\don} &=\big\langle\ket{\don}\bra{\don}\big\rangle_\qc,
\end{align}
%\end{subequations}
and \emph{vice versa} for $k_{\acc\to\don}^{(\mr s)}$ (where `s' stands for `side'). Here, $\Tp$ is the time required for the derivative $\dot K_{\don\acc}(t)$ to relax to a plateau after an initial transient on a shorter timescale.

Due to time-translational invariance, quantum correlation functions obey $K_{03}(t)= K_{30}(t)=\langle \sigma_3\rangle$, such that $K_{\acc\don}(t)=K_{\don\acc}(t)$ and hence detailed balance [\cref{eq:db_rates}] is obeyed. 
Spin-LSC and the optimized-sphere method do not fulfill this property: %they will in general have $K_{03}(t)\neq K_{30}(t)$ 
although $K_{30}(t)=\langle \sigma_3\rangle_\qc$, $K_{03}(t)$ is not correct at intermediate times in these methods (as demonstrated in the lower panel of the inset of \cref{fig:K33}), and thus within these approaches $K_{\acc\don}(t)\ne K_{\don\acc}(t)$, meaning that the forward and backward reaction rates differ and \cref{eq:db_rates} does not hold.
Ellipsoid mapping, on the other hand, being time-translationally invariant by construction, fulfills $K_{03}(t)=K_{30}(t)=\langle \sigma_3\rangle_\qc$ and $K_{\acc\don}(t)=K_{\don\acc}(t)$; hence it obeys detailed balance [\cref{eq:db_rates}]. %The local construction additionally conserves nuclear expectation values, which is not guaranteed by the global construction. % and thereby does obey detailed balance as defined in \cref{eq:db_rates}. 

%As a consequence of time-translational invariance, this condition is fulfilled by ellipsoid mapping, but not by the original spin-LSC or the optimized-sphere method.

Note, however, that there are alternative ways to compute the rate in practice. Using $\frac{\rm d}{\mathrm{d}t}\hat\sigma_3=2\Delta\hat\sigma_2$, one can write the rate constant as an integral of the
%The rates are calculated as a function of the electronic coupling constant $\Delta$, with two approaches: from the time integral of the \emph{electronic} 
electronic (`e') flux--flux correlation function, \cite{miller1983tromp} %according to
\begin{equation}\label{eq:k_e}
k^{(\mathrm e)}_{\don\to\acc} = \frac{\Delta^2}{P_\don}\int_0^{\Tp}\mr d t\,K_{22}(t),
\end{equation}
%\tcb{\begin{align}%keep this comment from here
%k^{(\mathrm e)}_{\don\to\acc}  &= -\frac {1}{4P_\don}\lim_{t\to \Tp}  \dot K_{33}(t) =- \frac {1}{4P_\don} \lim_{t\to \Tp} \langle \sigma_3\mc L\sigma_3(t)\rangle_{\qc}\\
%&= \frac {\Delta}{2P_\don}\lim_{t\to \Tp}  \langle \sigma_2\sigma_3(t)\rangle_{\qc}=\frac{\Delta^2}{P_\don}\int_0^{\Tp}\,\mr d\tau\,K_{33}(\tau).
%\end{align}
%Given that $\mc L\sigma_3 = 2\epsilon_{3jk}H_j\sigma_k=2\Delta\epsilon_{312}\sigma_2=2\Delta\sigma_2$,  $\sigma_3(\Tp)-\sigma_3(0) = 2\Delta\int_0^{\Tp}\,\mr d\tau\,\sigma_2(\tau)$, $\langle \sigma_3\sigma_2\rangle=0$.}\\ %to here
Another definition of the rate follows from using the nuclear side operator $\theta(y-y^\ddagger)$, where $y$ denotes a nuclear reaction coordinate and $y^\ddagger$ a dividing surface between donor and acceptor. This leads to the expression \cite{chandler1987greenbook}
%and from the plateau value of the \emph{nuclear} flux--side correlation function of the bath
\begin{equation}\label{eq:k_n}
k^{(\mathrm n)}_{\don\to\acc} = \frac{1}{P_{\rm d}^{(\mathrm n)}}\lim_{t\to\Tp} K^{(\mathrm n)}_{\mathrm{fs}}(t),
\end{equation}
% \tcb{From \cref{eq:k_da} we replace the donor by $\theta(y^\ddagger-y)$ and the acceptor by $\theta(y^\ddagger-y)$, to find 
% \begin{align*} %keep the comment
% k^{(\mathrm n)}_{\don\to\acc} &= \lim_{t\to\Tp} \frac{1}{P_\don} \frac{\mr d}{\mr d t}\langle \theta(y^\ddagger-y) \theta(y^\ddagger-y_t)\rangle \\
% &=  \lim_{t\to\Tp} -\frac{1}{P_\don} \langle\mc L \theta(y^\ddagger-y) \theta(y^\ddagger-y_t)\rangle \\
% &= \lim_{t\to\Tp} \frac{1}{P_\don} \langle \delta(y^\ddagger-y)\dot y \theta(y^\ddagger-y_t)\rangle
% \end{align*}}
where
\begin{equation}
K^{(\mathrm n)}_{\mathrm{fs}}(t)=\langle \dot{y}\delta(y-y^\ddagger)\theta(y_t-y^\ddagger)\rangle_\qc
\end{equation}
is the nuclear (`n') flux--side correlation function and $P_{\rm d}^{(\mathrm n)}=\langle \theta(y^\ddagger-y)\rangle_\qc$. %[$P_d$ needs to be updated too...]
% Here, $y$ denotes the nuclear reaction coordinate, as defined in \cref{eq:RCy}, and $y^\ddagger$ is the dividing surface on the phase space, which for $\varepsilon=0$ was fixed to $y^\ddagger=0$.
Both of these alternative definitions trivially obey the detailed-balance condition [\cref{eq:db_rates}] regardless of the underlying dynamics.

\begin{figure}
    \centering
    \includegraphics{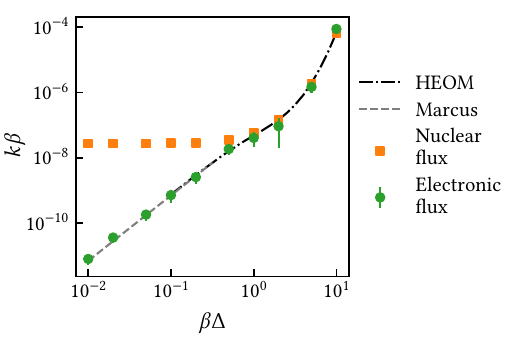}
    \caption{Reaction rates as a function of $\beta\Delta$ for a spin--boson model with $\Lambda=60$ and $\varepsilon=0$, comparing two different measurements using the local ellipsoid to numerically exact HEOM results from Ref.~\onlinecite{lawrence2019calculation}. If the rates are calculated from the plateau value of the \emph{nuclear} flux--side correlation function (orange squares) one obtains a trend akin to mean-field dynamics (valid only for large $\Delta$). Conversely, if the rates are calculated from the integral of the \emph{electronic} flux--flux correlation function, one obtains correct behaviour, including in the Marcus theory limit for small $\Delta$ (dashed line).}
    \label{fig:rate}
\end{figure}

One would like the rates calculated from the three approaches to be identical, but in the following we show that they are not.
In particular, although $k^{(\mathrm n)}_{\don\to\acc}=k^{(\mathrm s)}_{\don\to\acc}$ (because the rate is formally independent of the choice of dividing surface),\cite{Miller1993QTST} $k^{(\mathrm e)}_{\don\to\acc}$ differs from the others.
As a practical example, we study a symmetric spin--boson model ($\varepsilon=0$) with a Brownian oscillator spectral density, $J(\omega)=\frac{\Lambda}{2}\frac{\gamma\Omega^2\omega}{(\omega^2-\Omega^2)^2+\gamma^2\omega^2}$, where $\Lambda=60$ and $\beta\gamma=\beta\Omega=0.5$.\cite{lawrence2019calculation} For this model, we use $y$ as defined in \cref{eq:RCy} and $y^\ddagger=0$.
%; in the present case of normal regime, reactants and products are well defined by both a nuclear dividing surface and by electronic state operators. %in the Marcus inverted regime this would be a problem
%adiabatic regime
The results for $k^{(\mathrm n)}$ and $k^{(\mathrm e)}$ are shown as a function of $\Delta$ in \cref{fig:rate}.
We have verified numerically that $k^{(\mathrm s)}$ tend towards $k^{(\mathrm n)}$ across the whole range of parameters studied, but these results require a longer time to plateau and are not shown here.
%Numerically, we confirmed that $k^{(\mathrm n)}=k^{(\mathrm s)}$ across the whole range of parameters studied.

In the adiabatic limit (large $\Delta$), $k^{(\rm e)}$ is found to agree with $k^{(\rm n)}=k^{(\mathrm s)}$ and with the exact benchmark.
In this regime, the ellipsoid shrinks to a ``pancake''-like shape with all population on the lower adiabat leading to Born--Oppenheimer dynamics. % (but still with non-zero $\sigma_x$ and $\sigma_y$ components). % mapping approach is able to properly describe configurations in which population transfer between the two adiabats is hindered by a large energy difference.
%nonadiabatic regime
In the nonadiabatic limit (small $\Delta$), the electronic measure of the flux operator leads to rate constants in good agreement with Marcus theory, whereas the nuclear-flux approach fails to reproduce the correct result.
%.  However, the other two methods fail to reproduce the correct result.
Although it appears promising that the rate based on the electronic flux is accurate in this case, %in good agreement with Marcus theory,
it is clear that the underlying nuclear dynamics have the wrong physical behaviour.

Next, we present the reason for why there is a difference between the formulation based on the electronic fluxes ($k^{(\mr e)}$) and that based on the electronic side operators ($k^{(\mr s)}$). This can be understood by analyzing the time derivative of $\bm{\sigma}$ [\cref{eq:sigmadot}],
\begin{equation}
    \dot{\bm{\sigma}} = \g \dot{\bm{u}} + \dot{\g}\bm{u} + \dot{\bm{c}}.
\end{equation}
The first term describes rotation under a fixed ellipsoid shape, while the last two terms describe reshaping and translation of the ellipsoid. Due to the presence of these terms, the time derivative of the electronic side operator is not equal to the electronic flux operator, since $\dot{\sigma}_3\neq 2\Delta\sigma_2$. Hence, the time derivative $\dot{K}_{\rm da}$ is not equal to the electronic flux-side correlation function (and $\ddot{K}_{\rm da}$ is not equal to the electronic flux-flux correlation function).
The rate \cref{eq:k_n} obtained from the nuclear flux, however, is equivalent to \cref{eq:k_da} obtained from $\dot{K}_{\rm da}$ as the rate is formally independent of the choice of dividing surface within the extended mapping phase space.
%quantity $\dot{K}_{\rm da}$ does still equal $K_{\rm fs}^{(\rm n)}$, but
This explains why the rate based on the electronic flux is different from that based on the nuclear flux.

Finally, we consider why the rate based on the nuclear flux approach appears to be less accurate than that based on the electronic flux.
The nuclear flux reports directly on the dynamics of the nuclei, whereas the electronic flux is relatively insensitive to them.
It is therefore clear that the nuclei are not behaving as expected.
%However, this approach predicts incorrect rates
In the small-$\Delta$ limit, away from the diabatic crossing, the ellipsoid quickly reduces to the lower adiabat due to the large reorganization energy and in this way forces too much population transfer to occur.
This is a consequence of our choice of equations of motion, in which the ellipsoid parameters are determined solely by $x$ and are independent of the current value of $\bm{u}$.
%Correlation functions of the correct electronic flux operators, on the other hand, lead to rates that do recover Marcus theory in this limit.
% so that $\dot{K}_{\rm da}$ does not equal the electronic flux-side correlation function. In other words, the rate constants differ because the time derivative of electronic observables is not equal to the quantum derivative. The nuclear side-side agrees electronic side-side and nuclear
% Although we observe that \cref{eq:k_e} a time derivative %Whether or not the dynamics can be adapted to correct this issue is still an open question. 
%
% This is because the ellipsoid still reduces to the lower adiabat due to the large reorganization energy (except in a narrow region around the crossing). This reshaping of the ellipsoid causes population transfer in addition to what one would expect from diabatic coupling, so that the nuclear forces quickly reduce to the lower adiabat. However, the remaining extent in the $\sigma_x$ and $\sigma_y$ directions still allow measurements of their correlation functions, which give rates in agreement with Marcus theory in this limit. 
%
% Not fundamentally difference between nuclear and electronic flux, because electronic side-side agrees with nuclear flux-side (and side-side). but rather time derivative
%
As was stated in \cref{sec:local}, the equations of motions used here are not unique but could possibly be adapted to correct this issue in the future.
We also remark that other nonadiabatic trajectory methods such as surface hopping do not generally recover Marcus theory without additional decoherence corrections\cite{subotnik2016review}
and that this may provide inspiration for an alternative solution.

% Let us also remark that the present approach, given the assumption of classical nuclei, does not account for nuclear tunnelling or nuclear zero-point energy effects in the calculation of the rates. %not sure if these matter for the system studied here
% The construction of an ultimate rate theory from mapping method valid on all  regimes is still in progress, and we regard ellipsoid mapping as one step on the road to this goal. % direction.

% \subsection{Comparison with mean-field}
% In this subsection, we demonstrate that the ellipsoid mapping is significantly better behaved than a naive mean-field dynamics, which can also be constructed to preserve detailed balance for the down-side of destroying the nonadiabatic behaviour.
% In particular, it is known that mean-field methods do not correctly capture the $\Delta$ dependence of the rate for small $\Delta$.

% \red{don't need this if we show rates - but do say something about how our dynamics are better than moving on thermally weighted average surface}

\section{Outlook to nonadiabatic path-integral methods}
In this paper we have investigated equilibrium dynamics of a mixed quantum--classical system in which the nuclei are treated with classical statistics. This neglects nuclear zero-point energy and tunnelling, which may become important for large values of $\beta\omega_{\mathrm{c}}$. A range of methods have been proposed to include such effects in nonadiabatic dynamics by means of imaginary-time path-integral techniques. However, it has turned out to be difficult to make such methods obey detailed balance (according to the full quantum Boltzmann distribution), without breaking other important limits such as Rabi oscillations.\cite{althorpe2016non}
%Here, we briefly compare the methods developed in the present paper to a few path-integral methods
We can compare these to the methods developed in the present paper by considering the high-temperature (single-bead) limit, where nuclear quantum effects are unimportant.
% In this section we compare the spherical and ellipsoidal approaches studied in this work to other nonadiabatic techniques developed to study thermal equilibrium dynamics.

Perhaps the simplest way to formulate nonadiabatic path-integral dynamics is %to introduce resolutions of the identity over the nuclear space in the Boltzmann operator but to keep 
using a path integral over the nuclear degrees of freedom but
a discrete sum over the electronic degrees of freedom. This approach is sometimes referred to as mean-field RPMD\@.\cite{hele2011masters,saller2020chapter} In its single-bead limit, the method would evolve nuclei on the mean-field potential defined by $-\frac{1}{\beta} \log \Tr_\q[\e^{-\beta\hat{V}(x)}]$. Given that it does not provide any real-time information of electronic operators, it cannot capture the rate in the nonadiabatic limit (similar to those obtained with the nuclear flux operator in \cref{fig:rate}). 
A more elaborate mean-field method has been developed by Montoya-Castillo and Reichman,\cite{montoya-castillo2017PI} based on a path-integral simulation of the Wigner representation of the quantum Boltzmann distribution. Their approach propagates the trajectories with Ehrenfest dynamics, which does not preserve the equilibrium distribution or guarantee that correlation functions will relax to the correct long-time limits.
% Montoya-Castillo and Reichman have developed
% a path-integral approach to represent the Wigner representation of the quantum thermal density.\cite{montoya-castillo2017PI}
%In Ref.~\cite{montoya-castillo2017PI} has been developed 
% a quasiclassical approach to study thermal correlation functions in nonadiabatic systems. Their method is based on an imaginary-time path-integral simulation of the Wigner representation of the quantum thermal density.\cite{montoya-castillo2017PI} 
% The dynamics is then propagated with Ehrenfest mean-field theory. In contrast to ellipsoid mapping and the optimized sphere, this approach does not guarantee correlation functions to relax to the correct equilibrium long-time limits. [could we propagate it with the ellipsoid in the single-bead limit?]
%Although it includes nuclear quantum effects in the initial distribution, the nuclear dynamics are still defined in terms of fully classical equations of motion. In our approaches, we instead choose to consistently treat both the statistics and the dynamics of the bath classically, %, while restricting our system parameters to regimes in which the classical approximation for the nuclei is valid.
% finding that quantum statistics are not necessary to get reasonable results in the limit of small $\beta\omega$.

Another type of method represents the electronic as well as the nuclear degrees of freedom with path integrals using the MMST or spin mapping.\cite{richardson2013nrpmd,richardson2017vibronic,chowdhury2017coherent,bossion2021spin}
Since these approaches use a different Hamiltonian for dynamics and for statistics, they also do not preserve the equilibrium distribution even in the high-temperature limit.
% The fact that ellipsoid mapping preserves detailed balance represents an advantage compared to several other nonadiabatic RPMD methods based on mapping, which do not generally fulfill this property. \cite{richardson2013nrpmd,richardson2017vibronic,chowdhury2017coherent,bossion2021spin} We compare to the classical limit of such methods where the ring polymer localizes to a point.
Although one formulation of RPMD derived in Ref.~\onlinecite{ananth2013mvrpmd} does preserve the distribution, it does so at the expense of breaking ergodicity. In particular, the distribution is not positive definite and trajectories are unable to cross between positive and negative regions of the phase space. Importantly, the same method fails to reproduce the correct Rabi oscillations in the limit of an isolated subsystem. \cite{ananth2013mvrpmd} The distribution of ellipsoid mapping, on the other hand, is positive definite and conserved by the dynamics. Also, the ellipsoid method is capable of reproducing the correct Rabi oscillations in the limit of zero system--bath coupling.

Finally, an approach known as isomorphic-RPMD  \cite{miller3isomorphic2018} converts the path-integral problem into an isomorphic nonadiabatic system in which classical Boltzmann sampling over the effective Hamiltonian yields the correct statistics. The approach makes use of standard methods such as surface hopping\cite{tao2019isoRPSH,tao2020microcan} to solve the dynamics. This method would only obey detailed balance provided that the underlying classical nonadiabatic dynamics does, which in general is not the case for surface hopping.
In principle, one might attempt to combine the isomorphic-RPMD approach with ellipsoid mapping in order to fulfill detailed balance according to the full quantum Boltzmann distribution. However, even if the ellipsoid method could be improved to give consistent rate constants,
isomorphic-RPMD cannot capture tunnelling in the golden-rule limit regardless of the choice of the underlying dynamics.\cite{Lawrence2019isoRPMD}
%this would likely inherit the problems of both approaches.\cite{Lawrence2019isoRPMD}
% [In principle, one might attempt to combine the isomorphic-RPMD approach with ellipsoid mapping in order to fulfill detailed balance according to the full quantum Boltzmann distribution. However, %even in that case such a hybrid approach would suffer an important limitation. In particular 
% it has been shown that regardless of the choice of the underlying dynamics, the isomorphic-RPMD method does not reduce to instanton theory in the golden-rule limit, \cite{GoldenGreens} and thus cannot correctly capture tunnelling.\cite{Lawrence2019isoRPMD} %An older method presented in Ref.~\onlinecite{shushkov2012}, which combines surface hopping and RPMD, has only been tested in adiabatic limit, and is likewise unable to recover instanton theory.
% Hence, further work is still needed to find a path-integral version of the ellipsoid mapping.]
In conclusion, the search for a nonadiabatic version of RPMD is still an open question.

\section{Conclusions}
In this article, we have investigated the problem of detailed balance in mixed quantum--classical systems and explored new ways to calculate thermal equilibrium correlation functions. The direct extension of spin-LSC to thermal correlation functions as well as its optimized-sphere generalization are both found to break detailed balance in general. The global ellipsoid preserves electronic expectation values but does not sample the nuclear phase space correctly.
These issues are solved by the local ellipsoid, %introduced in this work (both in its local and global formulation) 
which rigorously conserves the quantum--classical equilibrium distribution and relaxes to the correct long-time limits, while predicting the correct Rabi oscillations in the limit of zero electron--nuclear coupling. These properties are not in general simultaneously fulfilled by any other classical-trajectory method that we are aware of. %Even the QCLE, which is often used to derive such methods, does not generally obey time-translational invariance of thermal distributions.\cite{nielsen2001QCbrackets}
By smoothly reducing $g_{ij}$ for energetically separated states, the local ellipsoid prevents trajectories from running off on inverted potentials and is therefore expected to avoid the difficulties for anharmonic systems experienced by previous mappings. % expected to be useful in the study of anharmonic systems.
%We also considered an alternative solution based on a sphere with an optimizated radius, which guarantee that selected correlation functions relax to the correct limit, even if it does not preserve the distribution.  
% The methods have been tested for a wide range of spin--boson systems at high temperature, including strongly-biased and weakly-coupled regimes, which are difficult to converge with available numerically exact methods. 

Although the local ellipsoid
is a solution to the long sought after method which simultaneously
obeys detailed balance as well as recovers Rabi oscillations,\cite{althorpe2016non} we found that these properties are not sufficient to give the correct relaxation timescales in the golden-rule limit. Specifically, we observed that $\dot{\sigma}_3\neq 2\Delta \sigma_2$, meaning that the time derivative of the electronic side operator is different from the electronic flux operator. We noted in \cref{sec:local} that the equations of motion were not unique, and it may be possible to adapt them to fulfill further properties than the ones listed in \cref{sec:ell_mixed}.
% They not unique in fulfilling the list of properties set up in \cref{sec:ell_mixed}, and could perhaps be adapted in the future to fulfill further chosen properties.
To outline how the dynamics may be improved, we point out that the ellipsoid approach can (like other mappings) be made formally exact by taking the phase-space representation of the full propagator. 
In the classical-nuclear limit, a reasonable starting point is the QCLE,
\begin{equation}
    \frac{\rd \hat{B}}{\rd t} = 2\epsilon_{ijk} H_i B_j \hat{\sigma}_k + \frac{p}{m}\frac{\partial \hat{B}}{\partial x} - \frac{1}{2}\left[\frac{\partial \hat{H}}{\partial x},\frac{\partial \hat{B}}{\partial p}\right]_+,
\end{equation}
which in the ellipsoid approach takes the form
\begin{equation}
    \frac{\rd B}{\rd t} = 2\epsilon_{ijk} H_i B_j (g_{kl}u_l+c_k) %\sigma_k 
    + \frac{p}{m}\frac{\partial B}{\partial x} - \frac{\partial H_\mu}{\partial x}\frac{\partial B_\mu}{\partial p}.
\end{equation}
To allow a solution in terms of independent trajectories, this needs to be approximated as
\begin{equation}
    \frac{\rd B}{\rd t} \approx \dot{u}_i\frac{\partial B}{\partial u_i} + \dot{x}\frac{\partial B}{\partial x} + \dot{p}\frac{\partial B}{\partial p}.
\end{equation}
There are many ways in which to perform this approximation. Our hope is that in the future an approximation can be found which fulfills  $\dot{\bm{\sigma}}=2\bm{H}\times\bm{\sigma}$, such that the time derivative of the electronic side operator becomes the electronic flux operator. This may require constraints or other specialized techniques in order to not break the condition $|\bm{u}|^2=1$. In any case, the ellipsoid approach presented here constitutes a rigorous framework on which further improvements can be based.
% Our particular choice for the equations of motion is clearly not a good approximation in all case.

% Constraint

An alternative development could be to process the current ellipsoid dynamics with the generalized quantum master equation (GQME), as has recently been done to improve the accuracy of nonequilibrium spin-mapping methods.\cite{amati2022gqme} The previous proof that GQME cannot improve the results of methods that obey detailed balance\cite{kelly2016master} does not hold in this case, because it would still correct for the issue that $\dot{\sigma}_3\neq 2\Delta \sigma_2$ in the current dynamics.

Although in the present work we focused on a two-state model for simplicity, we expect that the approach could be extended to more than two levels similarly to the original (spherical) spin mapping.\cite{runeson2020}
Finally, we restricted our analysis to purely classical nuclear dynamics neglecting nuclear zero-point energy and tunnelling, which is justified at high temperatures and low nuclear frequencies. 
It may be possible to include such effects in the future by combining it with path-integral methods such as RPMD. %Before going there, however, one should identify which properties have to be obeyed in order to capture golden-rule rates; otherwise such an extension will, like previous attempts, lead to incorrect timescales already in the classical-nuclear limit.
% we aim to include nuclear quantum effects in the approaches by combining the methods with the path-integral formalism. From this perspective, the methods presented in this work can be seen as one-bead limits of a more general nuclear ring polymer. 
%We have assumed that the dynamical system is mixing to relax to the long-time limit, which is typically valid for condensed-phase reactions but not applicable for scattering problems.
% Let us also remark that, even if all the methods discussed are trajectory-based, no physical meaning should be assigned to individual trajectories, but only to average properties as ensemble averages and time-correlation functions. %Our approaches are therefore suited to calculate exclusively average dynamical properties. 
% Also, the method relies on the assumption that the dynamics exhibits mixing behaviour, hence it is not suited to problems in which this property may not hold, such as the scattering problems described via Tully models. \cite{tully1990hopping}

% In future work, we aim to further improve the accuracy of the dynamics of ellipsoid mapping by exploring alternative equations of motion. In particular, we hope that this will resolve the current discrepancy between nuclear and electronic rates. %This would have positive implications in the predictions obtained from rate theory.
% Furthermore, we aim to include nuclear quantum effects in the approaches by combining the methods with the path-integral formalism. From this perspective, the methods presented in this work can be seen as one-bead limits of a more general nuclear ring polymer. 

\red{

}

\section*{Supplementary Material}
See the supplementary material for numerical data points of the results shown in \cref{fig:g_elements,fig:K33}.

\section*{Acknowledgements}
This project has received funding from European Union’s Horizon $2020$ under MCSA Grant No. $801459$, FP-RESOMUS. 
%\tcb{Other grants to mention?}

\section*{Author declarations}
\subsection*{Conflict of Interest}
The authors have no conflicts to disclose.

\subsection*{Author Contributions}
\textbf{Graziano Amati:} Formal analysis and data curation (lead of HEOM, thermal spin-LSC, optimized sphere in \cref{app:optr}, thermal averages), writing -- original draft (equal), writing -- review \& editing (equal).
\textbf{Johan E. Runeson:} Formal analysis and data curation (lead of optimized sphere in \cref{sec:optr}, SW generalization, ellipsoids, rates), writing -- original draft (equal), writing -- review \& editing (equal).
\textbf{Jeremy O. Richardson:} Formal analysis (supporting), writing -- review \& editing (equal).

\section*{Data availability}
%(use standard statement from JCP)
The data that support the findings of this study are available within the article and its supplementary material.

%%%%%%%%%%%%%%%%% APPENDIX %%%%%%%%%%%%%%%%%%%

\appendix
\section{Principal-axis basis}\label{app:principal}
In this section, we show how to obtain the ellipsoid parameters $\g$, $\bm{c}$, and $\tilde{H}$ in the isolated subsystem and in the local construction for mixed quantum--classical systems. For simplicity, we assume that the Hamiltonian is chosen to be real (i.e., $H_2=0$).

Rotating to the principal-axis basis corresponds to diagonalizing the Hamiltonian, $\hat{H}=\hat{U}\hat{H}^{\rm p}\hat{U}^\dagger$, where 
\begin{equation}
 \hat{U}=\begin{pmatrix}
\cos\vartheta & -\sin\vartheta \\ \sin\vartheta & \cos\vartheta
\end{pmatrix}.
\end{equation}
% real symmetric matrix diagonalized by orthogonal matrix. we choose this to be a rotation
The diagonal elements of $\hat{H}^\pr$ are the adiabatic energies, sorted in decreasing order if the $z^\pr$-axis is chosen to be aligned with the field, as in \cref{fig:principal}.  In this orientation,
% orientation in \cref{fig:principal}, 
$H_3^\pr=\frac{1}{2}([\hat{H}^\pr]_{11}-[\hat{H}^\pr]_{22})$ is positive.
%\tcb{\begin{equation*} %keep commented equation
%    \hat H^\pr = \hat\sigma_0 \frac 12 ([H^\pr]_{11}+[H^\pr]_{22})+ \hat\sigma_3 ([H^\pr]_{11}-[H^\pr]_{22})
%\end{equation*}}
%, so that 
% If the axes are oriented such that $H_3^{\rm p}\equiv \frac{1}{2}\Tr_\q[\hat{\sigma}_3\hat{H}^\pr] \geq 0$ (as in \cref{fig:principal}), then the diagonal elements of $\hat{H}^\pr$ are sorted in decreasing order, so that the principal-axis basis is the same as the adiabatic basis up to a reordering of the states. 
However, we shall present the following solution so that it is valid for both positive and negative $H_3^\pr$. 
%Our calculations were carried out in the original (diabatic) basis, but do give the same result if carried out in the adiabatic basis.

We define the rotation matrix $\mat{R}$ such that $\hat{U}^\dagger\hat{\sigma}_i\hat{U}=R_{ij}\hat{\sigma}_j$. Hence $H_i^{\rm p}%=(0,0,H^{\rm p}_3)
= H_j R_{ji}$ 
\tcb{%keep comment
\begin{align*}
&\hat U^\dagger \hat H\hat U = \hat H^\pr, \hspace{5mm}H_i R_{ij}\hat\sigma_j=H^\pr_j\hat\sigma_j, \hspace{5mm} R^TH=H^\pr\\
& H g u = R H^\pr (R g^\pr R^T) R u^\pr = H^\pr g^\pr u^\pr
\end{align*}
}
% (R^\intercal)_{ij}H_j$. 
and $R_{ij}= \thalf\Tr_\q[\hat{U}^\dagger\hat{\sigma}_i\hat{U}\hat{\sigma}_j]$, or more explicitly
%Symmetry implies that lower right $3\times 3$ block is symmetric $\implies$ can find diagonal basis where
\begin{equation}
 \qquad \mat{R} = \begin{pmatrix}
%1 & 0 & 0 & 0 \\ 
 \cos 2\vartheta & 0 & -\sin 2\vartheta \\
 0 & 1 & 0 \\
 \sin 2 \vartheta & 0 & \cos 2\vartheta
\end{pmatrix}.
\end{equation}
Given that scalar quantities are basis-independent, $\bm{H}^\intercal\g \bm{u}=(\bm{H}^\pr)^\intercal \g^\pr \bm{u}^\pr$ and $\bm{c}
\cdot\bm{u}=\bm{c}^\pr\cdot\bm{u}^\pr$, which is fulfilled by the transformation rules % the mapping parameters transform as 
$\bm{u}=\mat{R}^\intercal\bm{u}^\pr$, $\g=\mat{R}^\intercal \g^{\rm p} \mat{R}$,
$\bm{c}=\mat{R}^\intercal \bm{c}^\pr$, and $\tilde{H}=\tilde{H}^{\rm p}$.
% and $g^{\rm p}$ is given by Eq. XX.

In the principal-axis basis, the only non-zero expectation value is $\langle \hat{\sigma}_3\rangle_\q=-\tanh \zeta$. The corresponding phase-space average is
\begin{equation}
    \langle \sigma_3\rangle_\c = \frac{\int \rd\bm{u}\e^{-\beta H_3^\pr g_{33}^\pr u_3} (g_{33}^\pr u_3+c_3^\pr) }{\int \rd\bm{u}\e^{-\beta H_3^\pr g_{33}^\pr u_3}}  = -g_{33}^\pr L(g_{33}^\pr\zeta)+c_3^\pr,
\end{equation}
where $L(x)=\coth x - \frac{1}{x}$ is the Langevin function.
Equating the expectation values leads to an expression for $c_3^\pr$ in terms of $g_{33}^\pr$,
\begin{equation}
    c_3^\pr  = g_{33}^\pr L(g_{33}^\pr\zeta) -\tanh \zeta.
\end{equation}
Note that reversing the sign of $\zeta$ (i.e., flipping the direction of the principal $z$ axis) also reverses the sign of $c_3^\pr$ (assuming that $g_{33}^\pr>0$).

Next, we match the zero-time correlations. Recall that there is a circular symmetry about the field, so that $g_{11}^\pr=g_{22}^\pr$. By evaluating the traces and integrals analytically, the equation $\langle \hat{\sigma}_1,\hat{\sigma}_1\rangle_\q = \langle \sigma_1,\sigma_1\rangle_\c$ reduces to
\begin{equation}
\frac{\tanh\zeta}{\zeta}
%    \frac{\int \rd\bm{u}\e^{-\beta H_3^\p g_{33}^\pr u_3} g_{11}^\pr u_3 g_{11}^\pr u_3 }{\int \rd\bm{u}\e^{-\beta H_3^\pr g_{33}^\pr u_3}} 
= (g_{11}^\pr)^2 \frac{g_{33}^\pr \zeta\coth (g_{33}^\pr \zeta)-1}{(g_{33}^\pr \zeta)^2},
\end{equation}
which means that $g_{11}^\pr$ can also be expressed in terms of $g_{33}^\pr$.
The only remaining unknown is therefore $g_{33}^\pr$, which we obtain by solving
 $\langle \hat{\sigma}_3,\hat{\sigma}_3\rangle_\q =\langle \sigma_3,\sigma_3\rangle_\c$, or
\begin{equation}
\sech^2\zeta = \frac{1}{\zeta^2}-(g_{33}^\pr)^2\csch^2(g_{33}^\pr \zeta).
\end{equation}
This is a single transcendental equation which we solve numerically with a one-dimensional root search. It has a unique positive solution which is shown in \cref{fig:g_elements}. 

What remains is to determine $\tilde{H}^\pr =\tilde{H}$. This is done by matching the partition functions
\begin{subequations}
\begin{align}
    Z_\q &= 2\e^{-\beta H_0} \cosh \zeta, \\
    Z_\c &= 2 \e^{-c_3^\pr \zeta-\beta (H_0+\tilde{H})} \frac{\sinh(g_{33}^\pr \zeta)}{g_{33}^\pr \zeta},
\end{align}
\end{subequations}
so that
\begin{equation}
    \frac{\tilde{H}}{H_3^\pr} = -c_3^\pr - \frac{1}{\zeta}\log\frac{g_{33}^\pr\zeta \cosh\zeta}{\sinh(g_{33}^\pr\zeta)}. 
\end{equation}
The right-hand side is positive for $\zeta>0$ and negative for $\zeta<0$.
% Note that reversing the sign of $\zeta$ also reverses the sign of $\tilde{H}/H_3^\pr$, and 
Because $\zeta$ and $H_3^\pr$ have the same sign, it follows that $\tilde{H}$ is always positive.

\section{Calculation of quantum--classical thermal averages}\label{app:QC_ave}
In this appendix we discuss how to efficiently calculate the quantum--classical thermal averages on the left-hand side of \cref{eq:QCmom_problem1,eq:QCmom_problem2}, for the specific case of the spin--boson model.

To reduce the dimensionality of the phase-space averages, we note from \cref{H_SB} that the system and the bath are coupled only via a scalar nuclear reaction coordinate $y$ as defined  in \cref{eq:RCy}.
All other $F-1$ nuclear degrees of freedom can be identified as a secondary bath, which can be analytically integrated out in the calculation of static correlation functions. 

The bath and system--bath coupling terms in the spin--boson Hamiltonian are rewritten as a function of $y$ as \cite{Thoss2001hybrid, wang2017}
\begin{subequations}
\begin{align}\label{eq:H_RC}
H_\B &= \thalf\sum_{\alpha=1}^{F-1}\left[\tilde p^2_\alpha +\tilde\omega_\alpha^2\left(\tilde x_\alpha-\frac{\tilde c_\alpha y}{\tilde\omega_\alpha^2}\right)^2\right],\\
\hat H_\SB &= \kappa y\hat \sigma_3,
\end{align}
\end{subequations}
where 
\begin{equation}\label{def_kappa_Omega}
\kappa = \left(\sum_{\alpha=1}^F c_\alpha^2\right)^{1/2},\hspace{10mm}\Omega^2 =  \kappa^2\left(\sum_{\alpha=1}^F \frac{c_\alpha^2}{\omega_\alpha^2}\right)^{-1}.
\end{equation}
Here, $p_y$ is the momentum conjugate to $y$, while all variables related to the secondary bath are denoted with a tilde. The coupling constants $\tilde c_\alpha$ and frequencies $\tilde \omega_\alpha$ can be calculated from the relation between the spectral densities of the primary and secondary bath. \cite{garg1985} The quantum--classical partition function in the new variables becomes
\begin{align}
Z_\Q &= \int\mathrm d \tilde x\mathrm d \tilde p \mathrm dy \mathrm d p_y\,  \prod_{\alpha=1}^{F-1}\exp\left\{-\frac\beta 2\left[\tilde p^2_\alpha +\tilde\omega_\alpha^2\left(\tilde x_\alpha-\frac{\tilde c_\alpha y}{\tilde\omega_\alpha^2}\right)^2\right]\right\}\nn\\
&\times \exp\left\{-\beta\left(\frac{p_y^2}2+\Omega^2 \frac{y^2}{2 \kappa^2}\right)\right\} \Tr_\q [\e^{-\beta (\hat H_\S + \kappa y \hat \sigma_3)}], \label{eq:Z_Q_RC}
\end{align}
and analogous expressions can be written for $\langle \hat \sigma_i\rangle_\qc$ and $\langle \hat \sigma_i,\hat\sigma_j\rangle_\qc$. The reduced partition function
\begin{equation} \label{eq:tZ_Q_RC}
\tilde Z_\Q = \int\mathrm d y\,\e^{-\beta \Omega^2 \frac{(y-\varepsilon)^2}{2 \kappa^2}} \Tr_\q[\e^{-\beta (\Delta\hat \sigma_1+  y\hat \sigma_3)}]
\end{equation}
is obtained after integrating out in \cref{eq:Z_Q_RC} all nuclear variables apart from the reaction coordinate $y$. An unimportant prefactor from the integration of the secondary bath is omitted in \cref{eq:tZ_Q_RC}, as it cancels out in the normalization of phase-space averages of the electronic operators we are interested in. Also, by defining
\begin{equation}\label{eq:rho_y}
\hat{\rho}(y) = \e^{-\beta\Omega^2 \frac{(y-\varepsilon)^2}{2\kappa^2}} \e^{-\beta(\Delta \hat{\sigma}_1 + y \hat{\sigma}_3)},
\end{equation}
we can rewrite exact quantum--classical averages as
\begin{subequations}
\begin{align}
\tilde{Z}_\qc &= \int \rd y\,\Tr_\q[\hat{\rho}(y)],\\
\langle \hat{\sigma}_i \rangle_\Q &= \frac 1 {\tilde{Z}_\qc}\int \rd y\,\Tr_\q[\hat{\rho}(y)\hat{\sigma}_i],\label{sigma_i_RC} \\
\langle \hat{\sigma}_i, \hat{\sigma}_j \rangle_\Q &= \frac 1 {\tilde{Z}_\qc}\int \rd y \int_0^1 \rd\lambda \,\Tr_\q[\hat{\rho}(y)^{1-\lambda} \hat{\sigma}_i \hat{\rho}(y)^\lambda \hat{\sigma}_j]. \label{sigma_ij_RC}
\end{align}
\end{subequations}
The trace in the above expressions is easily evaluated and the integral over $\lambda$ can be performed analytically. Finally, the numerical solution of the integral over $y$ can be calculated via one-dimensional quadrature.

\section{Spherical spin mapping for thermal correlation function}
In this appendix we discuss two spherical mapping approaches suited to the study of thermal correlation functions. A comparison between these two methods and ellipsoid mapping is discussed in \cref{sec:results}.

\subsection{Spin-LSC}\label{app:SM_thermal}
We introduce here a formulation of spin-LSC \cite{spinmap, runeson2020} suited to the study of thermal correlation functions. The approach is based on the notion [already introduced in \cref{eq:TrS_mn}] that the electronic trace of the product of two electronic operators $\hat A$ and $\hat B$ can be represented as an integral over a spherical phase space
\begin{subequations}\label{eq:LSC}
\begin{equation}\label{eq:genSW}
\tr_\q[\hat A\hat B] = \int \mr d \bm{u}\, A_{\W}(\bm{u})B_{\W}(\bm{u}),  
\end{equation}
where
\begin{align}
A_{\W}(\bm{u}) &= \tr_\q[\hat A \hat{w}_{\W}(\bm{u})],\\
B_{\W}(\bm{u}) &= \tr_\q[\hat B \hat{w}_
\W(\bm{u})],\\
\hat w_\W(\bm{u}) &= \tfrac{1}{2} \hat{\mc I} + g_\W \hat \sigma_i u_i, 
\end{align}
\end{subequations}
and $g_\W=\sqrt{3}$ in standard spin-LSC \cite{spinmap}. %; also, $\mr d \bm{u}$ is defined as in \cref{eq:TrS_mn}.
Based on the mapping prescription in \cref{eq:LSC}, we define thermal expectation values and correlation functions within spin-LSC by
\begin{subequations}\label{eq:SM_i_ij}
\begin{align}
\langle \sigma_j(t)\rangle^{\W} &= \frac 1{Z_\Q} \int \mr d x\mr d p\,\e^{-\beta H_\B}\int \mr d \bm{u}\, \mc T_0^{\W}(x,\bm{u}) \sigma_j^{\W}(\bm{u}_t), \label{eq:SM_i}\\
K^{\W}_{ij}(t) &= \frac 1 {Z_\Q} \int \mr d x\mr d p\,\e^{-\beta H_\B}\int \mr d \bm{u}\, \mc T_i^{\W}(x,\bm{u}) \sigma_j^{\W}(\bm{u}_t),\label{eq:SM_ij}
\end{align}
\end{subequations}
% \begin{subequations}\label{eq:corrW}
% \begin{align}
% \langle \sigma_j\rangle^{\W} &= \frac 1 {Z_\Q} \int \mr d x\mr d p\,\e^{-\beta H_\B}\int \mr d \bm{u}\, \mc T_0^{\W}(\bm{u}, x) \sigma_j^{\W}(\bm{u}), \label{eq:sigmai_r}\\
% \langle  \sigma_i \sigma_j\rangle^{\W} &= \frac 1 {Z_\Q} \int \mr d x\mr d p\,\e^{-\beta H_\B}\int \mr d \bm{u}\, \mc T_i^{\W}(\bm{u}, x) \sigma_j^{\W}(\bm{u}),\label{eq:Kij_r}
% \end{align}
% \end{subequations}
where 
\begin{subequations}
\begin{align}
&\mc T_\mu^{\W}(x,\bm{u}) = \tr_\q[\hat{\mc T}_\mu(x)\hat w_\W(\bm{u})]\label{eq:T_mu_map}, \\
&\hat{\mc T}_\mu(x) = \frac 1 \beta \int_0^\beta\mr d \lambda\,\e^{-(\beta-\lambda)\left(\hat H_\S+\hat H_\SB(x)\right)}\hat\sigma_\mu \e^{-\lambda \left(\hat H_\S+\hat H_\SB(x)\right)}, \label{eq:T_mu_qm}
\end{align}
\end{subequations}
and $\sigma_i^\W(\bm u)$ has been defined in \cref{eq:oldmap}.
At $t=0$, $\langle \sigma_j(0)\rangle^\W$ and $K_{ij}^\W(0)$ %\cref{eq:sigmai_r} and \cref{eq:Kij_r} 
are equal by construction to their quantum--classical correspondents $\langle \hat \sigma_i\rangle_\qc$ and $\langle \hat\sigma_i,\hat \sigma_j\rangle_\qc$, respectively. 
When the variables $(x,p,\bm{u})$ are evolved under the Hamiltonian ${H_{\W}(x,p,\bm{u}) = \Tr_\q[\hat H\hat w_\W(\bm{u})]}$, the time evolution of $K_{ij}^\W(t)$ is an approximation to the exact dynamics of $K_{ij}(t)$ analogous to LSC-IVR. 

To evaluate \cref{eq:SM_i_ij}, we sample $(x,p)$ from the uncoupled bath distribution 
\begin{equation}
\hat\rho_\B = \frac{\e^{-\beta H_\B}}{Z_\B}, \hspace{10mm}
Z_\B =\int\mr d x\mr d p\, \e^{-\beta H_\B},
\end{equation}
and $\bm{u}$ uniformly from the unit sphere.
Since this initial distribution is effectively out of equilibrium [like \cref{eq:trAB}], spin-LSC will not conserve expectation values in time or lead to the correct long-time limits.
If we assume that correlation functions decorrelate at long times, the expressions in \cref{eq:SM_i_ij} will instead approach
\begin{subequations}\label{eq:K_i_ij_SM}
\begin{align}
\lim_{t\to+\infty}\langle \sigma_j(t)\rangle^{\W}&=\lim_{t\to+\infty} \langle \mc T_0^{\W}\sigma_j^{\W}(t)\rangle_0 \nn\\
&=\langle \mathcal T_0^{\W}\rangle_0 \langle \sigma_j\rangle_{\rm eq}^{\W}= \langle \sigma_j\rangle_{\rm eq}^{\W}, \label{eq:SM_i_LT}\\
\lim_{t\to+\infty}K^{\W}_{ij}(t)&=\lim_{t\to+\infty} \langle \mc T_i^{\W}\sigma_j^{\W}(t)\rangle_0 \nn \\
&=\langle \mathcal T_i^{\W}\rangle_0 \langle \sigma_j\rangle_{\rm eq}^{\W}=\langle \hat \sigma_i\rangle_\qc \langle \sigma_j\rangle_{\rm eq}^{\W},\label{eq:SM_ij_LT}
\end{align}
\end{subequations}
where we defined expectation values with respect to the initial distribution,
\begin{equation}
    \langle A \rangle_0 = \frac{1}{Z_\qc} \int \rd x \rd p \, \e^{-\beta H_\b} \int \rd \bm{u} \, A(x,\bm{u}),
\end{equation}
with $Z_\qc$ as defined in \cref{eq:Zqc}, and with respect to the equilibrium distribution
\begin{subequations}\label{eq:sigma_ir}
\begin{align}
\langle  A \rangle^{\W}_{\rm eq} &= \frac{1}{Z^{\W}_{\rm eq}}\int\mathrm d x \mathrm d p \, \int \mathrm d \bm{u}\,\e^{-\beta H_{\W}(x,p,\bm{u})} A_{\W}(x,\bm{u}), \\
Z^{\W}_{\rm eq} &= \int \mathrm d x \mathrm d p\int\mathrm d \bm{u}\, \e^{-\beta H_{\W}(x,p,\bm{u})}.
\end{align}
\end{subequations}
Also, we used that
\begin{align}
\langle  {\mathcal T}_\mu^{\W} \rangle_0 &= \frac 1 {Z_\qc}\int \mr d x\mr d p\, \e^{-\beta H_\B}\int \mathrm d \bm{u} \, \mathcal T_\mu^{\W}(\bm{u},x) \nn \\
&= \frac 1 {Z_\qc} \int\mathrm d x \mathrm d p \,\e^{-\beta H_\B} \mathcal \Tr_\q[\hat{\mathcal T}_\mu(x)]=  \langle \hat \sigma_\mu \rangle_\Q. \label{Ti_ave}
\end{align}
In general, $\langle \sigma_j \rangle^{\W}_{\rm eq} \ne \langle \hat{\sigma}_j \rangle_{\qc}$, and thus \cref{eq:K_i_ij_SM} leads to the wrong long-time limits.

\subsection{Optimized sphere}\label{app:optr}
The mapping prescriptions in \cref{eq:LSC} can be generalized to \emph{any} radius $g_s$ according to
\begin{subequations}
\begin{equation}\label{eq:genSWs}
\tr_\q[\hat A\hat B] = \int \mr d \bm{u}\, A_{s}(\bm{u})B_{\bar s}(\bm{u}),  
\end{equation}
where
\begin{align}
A_{s}(\bm{u}) &=\tr_\q[\hat A \hat{w}_s(\bm{u})],\\
B_{\bar s}(\bm{u}) &= \tr_\q[\hat B \hat{w}_{\bar s}(\bm{u})],\\
\hat w_s(\bm{u}) &= \tfrac{1}{2} \hat{\mc I} + g_s \hat \sigma_i u_i, %\hspace{10mm}t=s,\bar s 
\end{align}
\end{subequations}
where $\hat w_{\bar s}$ is defined in the same way but with $g_{\bar s}=3/g_s$. We can use the additional free variable to
correct one of the equilibrium expectation values of spin-LSC\@. In this way, one can make the mapping approximations of both $K_{i j}(t)$ and $\langle \hat \sigma_j(t)\rangle_\qc$ relax to the correct long-time limits for a chosen value of $j$ (and all $i=1,2,3$). Here, the time dependence refers to evolution under $H_s$.
In particular, we look for a numerical solution of $g_s$  to the nonlinear equation
\begin{equation}\label{id_optz}
\langle \sigma_j\rangle^{(s)}_{\rm eq} \stackrel{!}{=} \langle \hat\sigma_j\rangle_\Q,
\end{equation}
where $\langle \sigma_j\rangle^{(s)}_{\rm eq}$ is defined as in \cref{eq:SM_i}, % \cref{eq:sigmai_r}, 
but with a radius $g_s$. %The condition \cref{id_optz} guarantees that the generalization to the optimized radius of \cref{eq:K_i_ij_SM} relax to the correct quantum--classical values.

In the case of the spin--boson model, \cref{id_optz} can be simplified by integrating out all degrees of freedom except for an electronic coordinate $z=\cos\theta$, defined such that $\sigma_3^s (z)= g_sz$. In the specific case of $j=3$  (with the aim of studying $K_{33}(t)$), we find
\begin{align}\label{eq:sigma_3_r}
\langle \sigma_3\rangle^{(s)}_{\rm eq} 
&= \frac{\int_{-1}^1\mathrm d z\, g_s z f(z)}{\int_{-1}^1\mathrm d z\, f(z)}, \\
f(z) &= I_0\left(-\beta g_s\Delta\sqrt{1-z^2}\right)\e^{\beta g_s z (\Lambda g_s z/4-\varepsilon)},
\end{align}
% \begin{align}\label{eq:sigma_3_r}
% \langle \sigma_3\rangle^{(s)}_{\rm eq} 
% &= \frac {g_s}{\tilde Z_{\rm eq}^{(3)}} \int_{-1}^1\mathrm d z\,  I_0\left(-\beta g_s\Delta\sqrt{1-z^2}\right)\nn\\
% &\quad \times\e^{\beta g_s z (\Lambda g_s z/2-\varepsilon)} z,\\
% \tilde Z_{\rm eq}^{(3)} &= \int_{-1}^1\mathrm d z\,  I_0\left(-\beta g_s\Delta\sqrt{1-z^2}\right)\e^{\beta g_s z (\Lambda g_s z/2-\varepsilon)},
% \end{align}
where we introduced the modified Bessel function of the first kind,\cite{bowman2012} $I_0(x)$,
and the reorganization energy $\Lambda =\sum_{\alpha=1}^F 2\frac{c_\alpha^2}{ m_\alpha\omega_\alpha^2}$. The $z$ integrals were evaluated numerically.

%\section{Benchmark thermal correlations from the hierarchical equation of motion}\label{app:HEOM}
\section{Benchmark Kubo-transformed thermal correlation functions}\label{app:HEOM}
To assess the accuracy of our quantum--classical calculations, we used a numerically exact solution of the HEOM \cite{tanimura2020} calculated with the open-source \lstinline{pyrho} package.\cite{berkelbach} Even though HEOM treats both the electronic and the nuclear degrees of freedom quantum--mechanically, the method is relevant for our quantum--classical framework by restricting to the high temperature regime.
The solution of HEOM needs to converge over two parameters: the number of Matsubara modes, $K$, and the truncation level, $L$. % necessary to accurately approximate the nuclear dynamics, and the number of $L$ levels at which the hierarchy of equations is truncated. 
%The number of Matsubara modes tends to increase the more quantum effects become relevant;
Given that we restrict our analysis to high temperatures, it sufficed for us to fix $K=0$. This regime corresponds to an exponentially fast decorrelation of the autocorrelation function of the nuclear reaction coordinate, as expected for a classical bath. \cite{tanimura2014}
For all the systems whose HEOM solution is shown in \cref{fig:K33}, the results converged with a truncation level between $L=13$ and $L=19$.
% Truncation of the number of the hierachy of the HEOM to values between $L=13$ and $L=19$ ensures a good convergence for all the systems whose HEOM solution is shown in \cref{fig:K33}. 
For two of the most ``extreme'' systems (marked with the label ``no HEOM'' in the figure) the solution of the HEOM failed to converge in the parameter $L$ before calculations became too demanding, hence we could not include benchmark results for those two cases. This issue stresses the importance of the development of efficient classical-trajectory techniques which can tackle systems inaccessible by numerically exact methods.

To calculate Kubo-transformed correlation functions with HEOM, we firstly prepared the system in an arbitrary out-of-equilibrium density
\begin{align}\label{eq:rho0}
\hat \rho_0&=\tfrac{1}{2} \hat\sigma_0\otimes \hat\rho_\B,
\end{align}
where $\hat\rho_\B$ denotes the quantum thermal distribution for the uncoupled bath. We then propagated the dynamics
for an equilibration time $t_{\mathrm{eq}}\gg 1$, chosen such that each expectation value plateaus to a constant,
\begin{equation}\label{eq:HEOM_Teq}
\langle \hat \sigma_j(t)\rangle^\b_{\q}= \langle \hat\sigma_j\rangle_\Q, \hspace{10mm}\forall \; t \ge t_{\mathrm{eq}}
\end{equation}
where $\langle\cdot \rangle_{q}^\b$ denotes an average over the initial distribution \cref{eq:rho0}, and the right-hand side corresponds to the expected quantum--classical equilibrium average as defined in \cref{eq:Aqc}. The condition \cref{eq:HEOM_Teq} can be seen as a consistency test on whether the assumption of classical nuclei is valid for a given system. % Since \cref{eq:HEOM_Teq}
%is expected to be valid in the limit of classical nuclear statistics and dynamics, . 
This means that after the time $t_{\rm eq}$, the initial density has thermalized to $\e^{-\ii\hat{H}t}\hat{\rho}_0\e^{\ii\hat{H}t}=\frac{1}{Z}\e^{-\beta\hat{H}}$. 

% Thermal correlation functions are initialized after propagating the initial density for the time interval $T_{\mathrm{eq}}$, as we can assume from the condition \cref{eq:HEOM_Teq} that $\e^{\mc Lt}\hat\rho_\B = \frac 1 {Z_\qc}\e^{-\beta \hat H}$ for $t\ge T_{\mathrm{eq}}$.
The output of the HEOM simulations is the standard thermal correlation function $C_{ij}(t)$ in \cref{eq:CAB}, which was then Kubo-transformed using the Fourier-space identity\cite{craig2004rpmd}
% \begin{equation}
% C_{ij}(t) = \frac 1 {Z_\qc}\tr_\qc[\e^{-\beta\hat H}\hat \sigma_i\hat \sigma_j(t)],
% \end{equation}
% which is conveted to its Kubo-transformed representation from the identity in Fourier space
\begin{equation}\label{eq:C_to_K}
\tilde K_{ij}(\omega) = \tilde C_{ij}(\omega) \frac{1-\e^{-\beta\omega}}{\beta\omega},
\end{equation}
where $\tilde f(\omega) = \int_{-\infty}^{+\infty}\mr d t \, \e^{-\ii\omega t}f(t)$. Finally, \cref{eq:C_to_K} was transformed back to time domain to get $K_{ij}(t)$.

%\red{say how you got kubo-transformed functions}

%See on laptop numerics/mathematica/RC_ave.nb
\clearpage

\bibliography{grazianorefs, johanrefs, morerefs} % use both reference files - but please only add new refs to biblio.bib 
\end{document}